\documentclass[twocolumn,
showpacs,
showkeys,
preprintnumbers,
nofootinbib,
superscriptaddress,
amsmath,
amssymb,
floatfix,
secnumarabic,
aps,
pra,
a4paper,
notitlepage,
final,
]{revtex4-1}%

\usepackage[colorlinks=true,urlcolor=blue]{hyperref}
\usepackage{graphicx}
\usepackage{epsfig}

\usepackage[pdf]{pstricks}
\newcommand{\beq}{\begin{equation}}
\newcommand{\eeq}{\end{equation}}
\newcommand{\beqar}{\begin{eqnarray}}
\newcommand{\eeqar}{\end{eqnarray}}
\newcommand{\bal}{\begin{aligned}}
\newcommand{\eal}{\end{aligned}}

\def\ham{\hbox{$\cal H$}}
\def\dalam{\hbox
{\vrule\vbox{\hrule\hbox to 1ex{ \hfill}\kern 1 ex\hrule}\vrule}}

\def\1/2{\hbox{$ {1 \over 2}$ }}
\def\tr{\hbox{Tr}}

\def\h{\hbar}
\def\i/h{{i \over \h}}

\def\inf{\infty}
 
\def\v{\vec}

\def\a{\alpha}  
\def\b{\beta}  
 
\def\g{\gamma}  
\def\d{\delta} \def\D{\Delta} 
  
\def\e{\epsilon} \def\E{\hbox{$\cal E $}}

\def\r{\rho} \def\vr{\varrho}

\def\p{\psi} \def\P{\Psi}
\def\bp{\bar \psi}

\def\m{\mu}
\def\n{\nu}

  \def\vk{\varkappa}

\def\z{\zeta}

\def\tt{\theta} \def\vt{\vartheta}

\def\<{\langle}
\def\>{\rangle}

\def\({\left(}
\def\[{\left[}
\def\){\right)}
\def\]{\right]}

\usepackage{subfigure}

\usepackage[notcite,notref,color]{showkeys}  %Р”Р»СЏ РѕС‚РѕР±СЂР°Р¶РµРЅРёСЏ СЃРѕРґРµСЂР¶РёРјРѕРіРѕ label{} Рє С„РѕСЂРјСѓР»Р°Рј, СЂРёСЃСѓРЅРєР°Рј, С‚Р°Р±Р»РёС†Р°Рј, Р»РёС‚РµСЂР°С‚СѓСЂРµ  %(С‡С‚РѕР±С‹ РЅРµ РїРµСЂРµРёРјРµРЅРѕРІС‹РІР°С‚СЊ С„РѕСЂРјСѓР»С‹, РїРµСЂРµРґ РѕС‚РїСЂР°РІРєРѕР№ РІ Р¶СѓСЂРЅР°Р» РїР°РєРµС‚ РѕС‚РєР»СЋС‡РёС‚СЊ)

\usepackage{tikz}
\usetikzlibrary{decorations.pathreplacing}
\usetikzlibrary{decorations.pathmorphing}    %Р”Р»СЏ СЂРёСЃРѕРІР°РЅРёСЏ РєРѕРЅС‚СѓСЂРѕРІ РёРЅС‚РµРіСЂРёСЂРѕРІР°РЅРёСЏ 2
\usepackage{multirow}
\usepackage{dcolumn}		%РІС‹СЂР°РІРЅРёРІР°РЅРёРµ РїРѕ С‚РѕС‡РєРµ РІ С‚Р°Р±Р»РёС†Р°С…, РёСЃРїРѕР»СЊР·РѕРІР°С‚СЊ d РІРјРµСЃС‚Рѕ c РІ СЃРїРµС†РёС„РёРєР°С†РёРё РєРѕР»РѕРЅРєРё
\newcolumntype{.}{D{.}{.}{-1}}
\newcolumntype{i}[1]{D{.}{.}{#1}}

\newcommand{\myfrac}[2]{{\ifmmode{}^{#1}\!/_{\!#2}\else${}^{#1}\!/_{\!#2}$\fi}}

\begin{document}
\sloppy

\title{Estimating the realistic start-up of spontaneous emission in the supercritical QED}

\author{A.Krasnov and K.Sveshnikov}
\email{k.sveshnikov@gmail.com}
\email{aa.krasnov@physics.msu.ru}
\affiliation{Department of Physics and
Institute of Theoretical Problems of MicroWorld, Moscow State
University, 119991, Leninsky Gory, Moscow, Russia}

%%remember to change the email and affiliation!!!!!

%Remember to change the date
\date{\today}

%%%%%%%%%%%%%%%%%%%

\begin{abstract}

The scenario of spontaneous positron emission, caused by the supercritical Coulomb source with charge $Z$ and size $R$, is explored in essentially non-perturbative approach  with emphasis on the vacuum energy $\E_{VP}$\footnote{Throughout the paper the abbreviation "VP"\, stands for "vacuum polarization"\,.}, considered as a function of $R$ with fixed $Z>Z_{cr,1}\simeq 170$.   It is shown that  in the supercritical region the behavior of  $\E_{VP}(R)$ turns out to be quite different in  dependence on the  value of the Coulomb charge $Z$. In particular, spontaneous emission becomes a visible effect starting only from $Z \sim 300$. With further growth of $Z$ its intensity increases very rapidly, which opens up the possibility for unambiguous detection.  For such $Z$ the appearing picture of spontaneous  emission turns out to be well-established and quite transparent in terms of the resonance decay combined with vacuum shells formation. The intimate, but crucial role of $\E_{VP}(R)$ in the  total energy balance in the system is outlined. The additional problems of spontaneous  emission, caused by the lepton  number conservation, are also discussed.
\end{abstract}


\maketitle

\section{Introduction}

 So far, the behavior of QED-vacuum exposed to a supercritical EM-source is subject to an active research~\cite{Rafelski2016,Kuleshov2015a,*Kuleshov2015b,*Godunov2017,Davydov2017,Sveshnikov2017,
Popov2018,*Novak2018,*Maltsev2018,Roenko2018,Maltsev2019,*Maltsev2020}.
Of the main interest is the assumption that  in such external fields there should take place  a deep vacuum state reconstruction, caused by discrete levels diving into the lower continuum and accompanied by such nontrivial effects as spontaneous positron emission combined with vacuum  shells formation (see e.g., Refs.~\cite{Greiner1985a,Plunien1986,Greiner2012,Ruffini2010,Rafelski2016} and citations therein). In 3+1 QED,  such effects are expected for extended Coulomb sources of nucleus size with charges $Z>Z_{cr,1} \simeq 170$, which are large enough for direct  observation and probably could be created in low energy heavy ion collisions  at new heavy ion facilities like  FAIR (Darmstadt), NICA  (Dubna), HIAF (Lanzhou)~\cite{FAIR2009,Ter2015,MA2017169}.

The problem of spontaneous positron emission in the supercritical QED has a long story, starting from the pioneering works of Frankfurt (W.Greiner, B.Mueller, J.Rafelsky, et al.) and Moscow (Ya.Zeldovich, S.Gerstein, V.Popov, et al.) groups (see, e.g., Refs.~\cite{Rafelski2016,Greiner1985a,Zeldovich1972} and citations therein) in the early seventies. However, in the subsequent investigations of supercritical heavy ion collisions at  GSI (Darmstadt, Germany), rechecked later by Argonne Lab. (USA), no evidence of the diving phenomenon was found ~\cite{Mueller1994}. The next generation of accelerator facilities is expected to drive these investigations to a new level~\cite{FAIR2009,Ter2015,MA2017169}. The novel experimental study requires an updated  theoretical analysis.

 In the present paper the non-perturbative VP-effects, caused by the quasi-static supercritical Coulomb source with $Z>Z_{cr,1}$ and size $R$, are explored in terms of VP-energy $\E_{VP}$. $\E_{VP}$ plays an essential role in  the region of super-criticality, especially for spontaneous positrons emission, since the latter should be provided solely by the VP-effects  without any other channels of energy transfer. In particular, it is indeed the decline of $\E_{VP}$, which provides the spontaneous positrons with corresponding energy for emission.
  Being considered as a function of $Z$, VP-energy reveals with growing $Z$ a pronounced decline into the negative range, accompanied with negative jumps, exactly equal to the electron rest mass, which occur each time when the  discrete level dives into the lower continuum~\cite{Grashin2022a}.  At the same time, being considered as a function of $R$ with fixed $Z$,  $\E_{VP}(R)$ simulates in a quite reasonable way the non-perturbative VP-effects in  slow heavy ion collisions. The most intriguing point here is that in the supercritical region the behavior of  $\E_{VP}(R)$ turns out to be quite different in dependence on the value of the Coulomb  charge $Z$. In particular, for $Z<250$ it grows with decreasing $R$ and so leaves no chances for spontaneous positrons. The start-up of  positron emission turns out to be not less than $Z \simeq 250-260$ (depending on the concrete model of the Coulomb source). Spontaneous emission becomes a visible effect starting from $Z \sim 300$. With further growth of $Z$ its intensity increases very rapidly, which opens up the possibility for unambiguous detection on the nuclear conversion pairs background. In view of recent attempts in this field of interest
~\cite{Rafelski2016,Kuleshov2015a,*Kuleshov2015b,*Godunov2017,Popov2018,*Novak2018,*Maltsev2018,
Roenko2018,Maltsev2019,*Maltsev2020,FAIR2009,Ter2015,MA2017169}, these circumstances require for a special study.

     Without any serious loss of generality, such a study can be performed within the  Dirac-Coulomb (DC) problem  with external spherically-symmetric Coulomb potential, created by a uniformly charged sphere
\beq
\label{1.5a}
V(r)=- Q\,\( {1 \over R}\, \tt(R-r)+ {1 \over r}\, \tt(r-R) \)  \ ,
\eeq
or  charged ball
\begin{multline}\label{1.6}
V(r)=- Q\,\( {3\, R^2 - r^2 \over 2\,R^3 }\, \tt(R-r) \ +  \right. \\ \left.  + \ {1 \over r}\, \tt(r-R) \)  \ .
\end{multline}
Here and henceforth
\beq
\label{1.5b}
Q=Z \a \ ,
\eeq
while the radius $R$ of the source varies in the range from
\beq
\label{1.8}
R_{min}(Z) \simeq 1.2\, (2.5\, Z)^{1/3} \ \hbox{fm} \ ,
\eeq
which roughly imitates  the size of a super-heavy nucleus with charge $Z$, up to certain $R_{max}$ of order of one electron Compton length, where the VP-effects are already small and show up as  $O(1/R)$-corrections.

The non-stationary approach, based on the time-dependent picture created by two  heavy ions, slowly moving along the classical Rutherford trajectories~\cite{Reinhardt1981,*Mueller1988,Ackad2008,Popov2018,*Novak2018,*Maltsev2018,
Maltsev2019,*Maltsev2020}, looks more attractive, since it imitates the realistic scenario of attaining the super-critical region in heavy ion collisions. At the same time,  the VP-effects at short internuclear distances achieved in the monopole approximation are in rather good agreement with exact two-center ones~\cite{Reinhardt1981,*Mueller1988,Tupitsyn2010,Maltsev2019,*Maltsev2020} and lie always in between those for the sphere and ball upon adjusting properly the coefficient in the relation (\ref{1.8}), because they are very sensitive to the latter. The main advantage of time-dependent approach is $ab \ initio$ description  of  pairs production  caused by the Coulomb excitations of nuclei, while in the adiabatic picture the latter should be considered  as an additional component. However, it will be argued below that the actual threshold for  vacuum positrons detection on the conversion pairs background turns out to be not less than $Z^\ast \simeq 250$, which lies beyond the existing nowadays  opportunities in heavy ion collisions.

It  should be specially noted that the parameter $Q$ plays actually the role of the effective coupling constant for the VP-effects under question. The size of the source  and its shape are also the additional input parameters, but their role in VP-effects is quite different from $Q$ and in some important questions, in particular, in the renormalization procedure, this difference must be clearly tracked. Furthermore, the difference between the charged sphere and the ball, which seems more preferable as a model of a super-heavy nucleus or heavy-ion cluster, in the VP-effects of  Coulomb super-criticality  is  small. It  shows up mainly in the ratio $ \E_{VP,ball}/\E_{VP,sphere} \simeq 6/5$ for the same $Z$ and $R$ with the only condition that $R$ must be close to $R_{min}(Z)$~\cite{Grashin2022a}. At the same time,   the charged sphere model allows for an almost completely analytical study of the problem, which has clear advantages in many positions. The ball model doesn't share such options, since  explicit solution of the DC problem in this case is absent and so one has to use from beginning the numerical methods or special approximations~\cite{Grashin2022a}.

As in basic works on this topic ~\cite{Wichmann1956,Gyulassy1975, McLerran1975a,*McLerran1975b,*McLerran1975c,Greiner1985a,Plunien1986,Greiner2012,Ruffini2010, Rafelski2016},  radiative corrections from virtual photons are neglected. Henceforth, if it is not stipulated separately, relativistic units  $\hbar=m_e=c=1$ and the standard representation of  Dirac matrices are used. Concrete calculations, illustrating the general picture, are performed for $\a=1/137.036$ by means of Computer Algebra Systems (such as Maple 21) to facilitate  the analytic calculations  and GNU Octave code for boosting the numerical work.

The paper is arranged as follows.  In Section II a consistent study of VP-density and the vacuum shells formation in the supercritical case is presented. In Section III we consider the evaluation of VP-energy in an essentially  non-perturbative approach with emphasis on renormalization and on the most effective methods of calculation. Section IV is devoted to presentation of results in the case of spherically-symmetric  Coulomb source (uniformly charged sphere and ball), the most important of which are shown in Figs.\,\ref{VP200-300(R)}-\ref{VP500(R)}. With account of these results, in Section V the question of what is wrong and what is possible in spontaneous emission is explored. Section VI is devoted to concluding remarks and general discussion.

\section{Vacuum shells formation}

The most efficient non-perturbative evaluation of the  VP-charge density $\vr_{VP}(\vec{r})$ is based on the  Wichmann and Kroll (WK) approach ~\cite{Wichmann1956,Gyulassy1975,Mohr1998}. The starting point of the latter is the vacuum value
\beq \label{3.1}
\vr_{VP}(\vec{r})=-\frac{|e|}{2}\(\sum\limits_{\e_{n}<\e_{F}} \p_{n}(\vec{r})^{\dagger}\p_{n}(\vec{r}) -  \sum\limits_{\e_{n}\geqslant \e_{F}} \p_{n}(\vec{r})^{\dagger}\p_{n}(\vec{r}) \) \ .
\eeq
 In (\ref{3.1}) $\e_F=-1$ is the Fermi level, which in such problems with strong Coulomb fields is chosen at the lower threshold, while $\e_{n}$ and $\p_n(\vec{r})$ are the eigenvalues and properly normalized set of  eigenfunctions of the corresponding DC problem. The expression (\ref{3.1}) for the  VP-charge density is a direct consequence of the well-known Schwinger prescription for the  fermionic current in terms of the fermion fields commutators
\beq
\label{3.1a}
 j_{\mu}(\v r, t)=-{|e| \over 2}\, \[ \bp(\v r, t)\,, \g_{\mu} \p(\v r, t)\] \ .
 \eeq

\begin{figure}
\center
\includegraphics[scale=0.20]{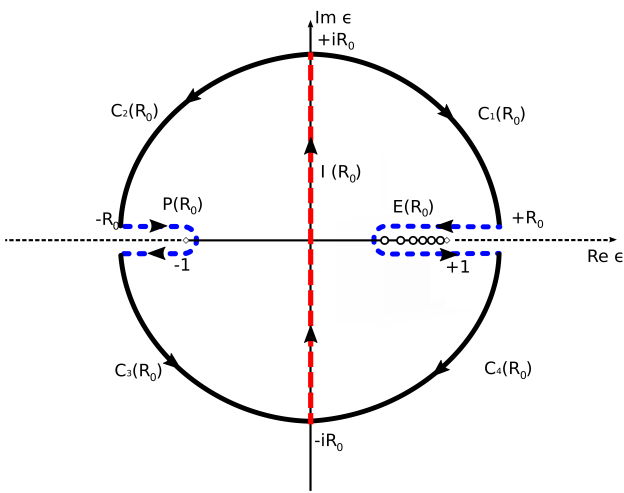} \\
\caption{\small (Color online) WK-contours in the complex energy plane, used for representation of the VP-charge density (\ref{3.1}) via contour integrals. The direction of contour integration is chosen in correspondence with (\ref{3.3}).}
\label{WK}
\end{figure}

The essence of the WK techniques is the representation of the density (\ref{3.1}) in terms of contour integrals on the first sheet of the Riemann energy plane, containing the trace of the Green function of  the corresponding DC problem. In our case, the  Green function is defined via equation
\beq
\label{3.2}
\[-i \v{\a}\,\v{\nabla}_r +\b +V(r) -\e \]G(\vec{r},\vec{r}\,' ;\e)  =\d(\vec{r}-\vec{r}\,' ) \ .
\eeq
The formal solution of (\ref{3.2}) reads
\beq
\label{3.3}
G(\vec{r},\vec{r}\ ';\e)=\sum\limits_{n}\frac{\p_{n}(\vec{r})\p_{n}(\vec{r}\ ')^{\dagger}}{\e_{n}-\e} \ .
\eeq
Following Ref.~\cite{Wichmann1956},  the density (\ref{3.1}) is expressed via the integrals along the contours  $P(R_0)$ and $E(R_0)$ on the first sheet of the complex energy surface  (Fig.\ref{WK})
\begin{multline}
\label{3.4}
\vr_{VP}(\vec{r}) = \\ = -\frac{|e|}{2} \lim_{R_0\rightarrow \infty}\( \frac{1}{2\pi i}\int\limits_{P(R_0)} \! d\e\, \mathrm{Tr}G(\vec{r},\vec{r'};\e)|_{\vec{r'}\rightarrow \v r} \ + \right. \\ \left.+ \ \frac{1}{2\pi i}\int\limits_{E(R_0)} \! d\e\, \mathrm{Tr}G(\vec{r},\vec{r'};\e)|_{\vec{r'}\rightarrow \v r} \) \ .
\end{multline}
Note that the Green function in this relation must be properly regularized to insure that the limit $\vec {r'} \to \v r$ exists and that the integrals over $d\e$ converge. This regularization is discussed below. On this stage, though, all expressions are to be understood to involve only regulated Green functions. One of the main consequences of the last convention is the uniform asymptotics of the integrands in (\ref{3.4}) on the large circle $|\e| \to \inf$ at least as $O(1/\e)$, which allows for deforming the contours  $P(R_0)$ and $E(R_0)$ to the imaginary axis segment $I(R_0)$ and taking the limit $R_0 \to \inf$, what gives
\begin{multline}
\label{3.4a}
\vr_{VP}(\vec{r}) = \\ =|e| \[ \sum \limits_{-1 \leqslant \e_n < 0} |\p_n (\v r)|^2 +{1 \over 2\pi}\,  \int\limits_{-\inf}^{\inf} \! dy\, \mathrm{Tr}G(\vec{r},\vec{r'};iy)|_{\vec{r'}\rightarrow \v r}\]  \ ,
\end{multline}
where $\{\p_n(\v r)\}$ are the normalized eigenfunctions of negative discrete levels with $-1 \leqslant \e_n <0$ and here and henceforth
\beq\label{3.4b}
\p_{n}(\vec{r})^{\dagger}\p_{n}(\vec{r}) \equiv |\p_n (\v r)|^2 \ .
\eeq
Proceeding further, the Green function (\ref{3.3}) is represented as a partial series over $k=\pm(j+1/2)$ ~\cite{Wichmann1956,Gyulassy1975}
\beq
\label{3.12}
\mathrm{Tr}G(\vec{r},\vec{r'};\e)|_{\vec{r'}\rightarrow \v r} = \sum\limits_k { |k| \over 2 \pi}\,\mathrm{Tr}G_k(r,r';\e)|_{r'\rightarrow r} \ ,
\eeq
where the radial Green function  $G_k(r,r';\e)$ is defined via
\beq
\label{3.13}
\ham_k(r)\,G_k(r,r';\e)={\d (r-r') \over r r'} \ ,
\eeq
while the  radial DC hamiltonian takes the form
\beq
\label{3.13a}
\ham_k(r)=  \begin{pmatrix} V(r)+1-\e & -{1 \over r}\,{d \over dr}\,r - {k \over r} \\ {1 \over r}\,{d \over dr}\,r - {k \over r} & V(r)-1-\e  \end{pmatrix}  \ .
\eeq
 For the partial terms of $\vr_{VP}(r)$ one obtains
\begin{multline}
\label{3.14}
\vr_{VP,k}(r)  = {|e| |k| \over 2 \pi} \[ \sum \limits_{-1 \leqslant \e_{n,k} < 0} |\p_{n,k} (r)|^2 + \right. \\ \left. + {1 \over 2\pi}\,  \int\limits_{-\inf}^{\inf} \! dy\, \mathrm{Tr}G_k(r,r';iy)|_{r'\rightarrow r}\]  \ ,
\end{multline}
where $\p_{n,k} (r)$ are the normalized radial wave functions with eigenvalues $k$  and $\e_{n,k}$  of the corresponding radial DC problem. From (\ref{3.14}) by means of symmetry relations for $G_k(r,r';\e)$ from Ref.~\cite{Gyulassy1975}   for the sum $\vr_{VP,|k|}(r)$ of two partial VP-densities with opposite signs of $k$  one finds
\begin{multline}
\label{3.21}
\vr_{VP,|k|}(r)  =  {|e| |k| \over 2 \pi}\, \Big\{\sum \limits_{k=\pm |k|} \sum \limits_{-1 \leqslant \e_{n,k} < 0}  |\p_{n,k} (r)|^2  + \\ + {2 \over \pi}\,  \int\limits_{0}^{\inf} \! dy\, \mathrm{Re}\[\mathrm{Tr}G_{k}(r,r';iy)\]|_{r'\rightarrow r} \Big\} \ ,
\end{multline}
which is  by construction real and odd in $Z$ (in accordance with the Furry theorem).

The general result, obtained in Ref.~\cite{Gyulassy1975} within the expansion of $\vr_{VP}(r)$ in powers of  $Q$ (but with fixed $R$\,!)
 \beq
\label{3.22}
\vr_{VP}(\v r)  =  \sum \limits_{n=odd}\,Q^n\,\vr_{VP}^{(n)}(\v r) \ ,
\eeq
is that all the divergencies of $\vr_{VP}(\v r)$ originate  from the fermionic loop with two external photon lines, whereas  the next-to-leading orders of expansion in $Q$ are already free from divergencies (see also Ref.~\cite{Mohr1998} and refs. therein). This statement is valid for 1+1 and 2+1 D always, and for a spherically-symmetric external potential in the three-dimensional case.  In the  non-perturbative approach this statement has been verified for 1+1 D in Refs.~\cite{Davydov2017,Sveshnikov2017} and for 2+1 D in Refs.~\cite{Davydov2018a,*Davydov2018b,Sveshnikov2019a,*Sveshnikov2019b} via direct calculation. The latter can be also extended for the present 3+1 D case\,\footnote{We drop here all the intermediate steps, required for the explicit construction of radial Green functions $G_{k}(r,r';\e)$ for the potentials (\ref{1.5a},\ref{1.6}) and justification of the limit $r' \to r$. For these steps one needs to deal with  explicit solutions of the DC problem and corresponding Wronskians, which takes a lot of space and so will be considered separately.}. It should be mentioned, however, that in  the three-dimensional case the spherical symmetry of the Coulomb potential is crucial, since the above-mentioned statement can be justified first within the partial expansion of $\vr_{VP}(r)$ for each partial term  $\vr_{VP,|k|}(r)$, but not for the whole series at once.

 This circumstance has been discussed earlier in Ref.~\cite{Gyulassy1975}, where it was shown that  the main reason for such difference is that the properties of the partial $G_k$  are much better than for the whole  series. First, the limit $r' \to r$ exists for $G_k\(r, r'; \e\)$, while the limit $\vec {r'} \to \v r$ does not exist for the total $ G\(\v r,\v r'; \e\)$.   At the same time, for the first-order (linear in $Q$)  VP-density $\vr^{(1)}_{VP,|k|}(r)$  different results are obtained from  (\ref{3.21}) if the limit $r' \to r$ and the contour integral are interchanged. Hence, it is indeed the first-order VP-density, which should be subject of renormalization via fermionic loop. To the contrary,  the direct calculation of the total $\vr_{VP}^{(3)}(\v r)$ contribution by means of the expression (\ref{3.4a}) suffers from an ambiguity associated with the interchange of the limit $\vec {r'} \to \v r$ and integration over imaginary axis, but the calculation of the contribution from each partial $\vr^{(3)}_{VP,|k|}(r)$ is free from such ambiguities. This study of   $\vr_{VP}^{(3)}$-contribution suggests that for bounded potentials, regularization of $\vr_{VP}^{(3)}(\v r)$ is achieved  by calculating $\vr_{VP}^{(3)}(\v r)$ as a sum over the partial  contributions $\vr^{(3)}_{VP,|k|}(r)$ without any other manipulations.

So the renormalization of the VP-density (\ref{3.1}) is actually the same for all the three spatial dimensions and implies the following procedure.
First,  the linear in the external field terms in the expression  (\ref{3.4a}) should be extracted and replaced further by the renormalized  first-order perturbative density $\vr^{PT}_{VP}(r)$, evaluated with the same $R$. For these purposes let us introduce  the component  $\vr^{(3+)}_{VP,|k|}(r)$ of partial VP-density, which is defined in the next  way
\begin{multline}\label{3.25}
\vr_{VP,|k|}^{(3+)}(r)= \frac{|e||k|}{2 \pi}\,\Big\{ \sum\limits_{k=\pm|k|}\sum\limits_{-1\leqslant \e_{n,k}<0}|\p_{n,k}(r)|^2 \ +  \\  + \ \frac{2}{\pi} \int\limits_{0}^{\infty}d y\,\mathrm{Re}\[\tr G_{k}(r,r;iy)- \tr G^{(1)}_{k}(r;i y)\] \Big\} \ ,
\end{multline}
where $G^{(1)}_{k}(r;i y)$ is the  linear in $Q$ component of the partial Green function $G_{k}(r,r;i y)$ and so coincides with the first term of the Born series
\beq \label{3.26}
G^{(1)}_{k}=G^{(0)}_{k} (-V) G^{(0)}_{k} \ ,
\eeq
where $G_{k}^{(0)}$ is the free radial Green function with the same  $k$ and $\e$.
By construction $\vr^{(3+)}_{VP,|k|}(r)$ contains only the odd powers of $Q$, starting from $n=3$, and so is free of divergencies. At the same time, it is responsible for all the nonlinear effects, which are caused by   discrete levels  diving into the lower continuum.

As  a result, the renormalized  VP-charge density $\vr^{ren}_{VP}(r)$ is defined by the following expression\,\footnote{The convergence of the partial series in (\ref{3.27}) is shown explicitly for the point source in the original work by Wichmann and Kroll~\cite{Wichmann1956}, while accounting for the finite size of the source is discussed in detail in Ref.~\cite{Gyulassy1975}. For the present case it follows from convergence of the partial series for  $\E^{ren}_{VP}$, which is explicitly shown in~\cite{Grashin2022a}.}
\beq\label{3.27}
\vr^{ren}_{VP}(r)=\vr_{VP}^{PT}(r)+\sum\limits_{k=1}^{\inf}\vr_{VP,|k|}^{(3+)}(r) \ .
\eeq
The first-order perturbative VP-density is obtained from the one-loop (Uehling) potential $A^{PT}_{VP,0}(\vec{r})$ in the next way ~\cite{Schwinger1949,*Schwinger1951,Itzykson1980, Greiner2012}
\beq
\label{2.2}
\vr^{PT}_{VP}(\vec{r})=-\frac{1}{4 \pi} \D A^{PT}_{VP,0}(\vec{r}) \ ,
\eeq
where
\beq\begin{gathered}
\label{2.3}
A^{PT}_{VP,0}(\vec{r})=\frac{1}{(2 \pi)^3} \int d \v q\,\, \mathrm{e}^{i \vec{q} \vec{r}}\, \Pi_{R}(-{\v q}^2)\,\widetilde{A}_{0}(\vec{q}) \ , \\
\widetilde{A}_{0}(\vec{q})=\int d \v r'\,\mathrm{e}^{-i \vec{q} \vec{r\,}' }\,A^{ext}_{0}(\vec{r}\,' ) \ .
\end{gathered}\eeq
The polarization function  $\Pi_R(q^2)$, which enters eq. (\ref{2.3}), is defined via general relation $\Pi_R^{\m\n}(q)=\(q^\m q^\n - g^{\m\n}q^2\)\Pi_R(q^2)$ and so is dimensionless. In the adiabatic case under consideration  $q^0=0$ and $\Pi_R(-{\v q}^2)$ takes the  form
\begin{multline}
\label{2.4}
\Pi_R(-{\v q}^2) = \\ = {2 \a \over \pi}\, \int \limits_0^1 \! d\b\,\b(1-\b)\,\ln\[1+\b(1-\b)\,{{\v q}^2 \over m^2-i\e}\] = \\ = {\a \over \pi}\, S\(|\v q|/m \) \ ,
\end{multline}
where
\begin{multline}\label{2.4a}
S(x)= -5/9 + 4/3 x^2 + (x^2- 2)\, \sqrt{x^2+4} \ \times \\ \times \ \ln \[ \(\sqrt{x^2+4}+x\) \Big/ \(\sqrt{x^2+4}-x\) \]/3x^3  \ ,
\end{multline}
with  the following IR-asymptotics
\beq
\label{2.4b}
S(x) \to x^2/15 -x^4/140  + O(x^6) \ , \quad x \to 0 \ .
\eeq

 The important feature of the renormalization procedure (\ref{3.27}) is that it provides vanishing of the  total renormalized VP-charge
\beq
\label{3.27a}
Q^{ren}_{VP}=\int\limits d {\v r}\,\vr^{ren}_{VP}(\v r) \ ,
\eeq
in the subcritical region $Z<Z_{cr,1}$. Actually, it is an  indication in favor of the assumption, that in presence of the external field, uniformly vanishing at the spatial infinity and without any specific boundary conditions and/or nontrivial topology of the field manifold, one should expect that in the region  $Z<Z_{cr,1}$ the correctly renormalized  VP-charge must vanish, while the VP-effects are able to produce only small distortions of its spatial distribution \cite{Greiner2012,Mohr1998}. It should be noted, however, that this is not a theorem, but just a plausible assumption, which in any concrete case should be verified via direct calculation.

In the present case the validity of this assumption can be shown in the following way. First, one finds that
\beq
\label{3.27b}
Q_{VP}^{PT} \equiv 0  \ .
\eeq
The relation (\ref{3.27b}) is nothing else but the direct consequence of the general QED-renormalization condition ${\Pi_R} (q^2) \sim  q^2$ for $q \to 0$, which is exactly confirmed by the asymptotics (\ref{2.4b}). More concretely, in the momentum space (up to factors like $2 \pi$ and general signs) the static equation for the potential $A_0(\v q)$, generated by the external charge density $\vr_{ext}(\v q)$, reads
\beq \label{2.5}
\({\v q}^2-{\tilde \Pi_R}(-{\v q}^2)\)A_0(\v q)=\vr_{ext}(\v q) \ ,
\eeq
where ${\tilde \Pi_R}(q^2)$ is the polarization operator in the standard form with dimension $[q^2]$, which is introduced via relation $\Pi_R^{\m\n}(q)=\(g^{\m\n}-q^\m q^\n/q^2\){\tilde \Pi_R}(q^2)$. Within PT one should propose that ${\tilde \Pi_R}(q)\ll q^2$, therefore  the potential $A_0(\v q)$ can be represented by the series
\beq \label{2.6}
 A_0(\v q)=A^{(0)}(\v q)+A^{(1)}(\v q)+\ldots.
 \eeq
 The classical part of the potential $A^{(0)}(\v q)$ is obtained from the equation
\beq \label{2.7}
{\v q}^2 A^{(0)}(\v q)=\vr_{ext}(\v q) \ ,
\eeq
and determines the external potential $A_0^{ext}(\v r)$, while its first quantum correction --- from the following one
\beq \label{2.8}
{\v q}^2 A^{(1)}(\v q)={\tilde \Pi_R}(-{\v q}^2)A^{(0)}(\v q)  \ .
\eeq
Up to the factor  $1/ 4 \pi$ the r.h.s. in (\ref{2.8}) is indeed the perturbative VP-density $\vr^{PT}_{VP}(\v q)$. Then in the coordinate representation (up to multipliers like $2\pi$) one gets
\begin{multline} \label{2.9}
\vr^{PT}_{VP}(\v r)=  \int \! d \v q \ \mathrm{e}^{i\v q \v r}\,{\tilde \Pi_R}(-{\v q}^2)\, A^{(0)}(\v q)= \ \\ \ = \int \! d \v r'\, {\tilde \Pi_R}(\v r-\v r') A^{ext}_{0}(\v r') \ .
\end{multline}
Upon integrating this expression over whole space  one obtains the total perturbative VP-charge $Q^{PT}_{VP}$ in the form
\beq \label{2.10}
Q^{PT}_{VP}=\int \! d \v q \ \d(\v q) \ {\tilde \Pi_R}(-{\v q}^2)\, A^{(0)}(\v q)   \ .
\eeq
Proceeding further by means of the  renormalization condition for ${\tilde \Pi_R}(q^2)$, one finds from  (\ref{2.10}) that  $Q^{PT}_{VP}=0$ under assumption, that in the momentum space  the singularity of the external potential $A^{ext}_{0}(\v q)$ for $|\v q| \to 0$ is weaker, than $O\(1/|\v q|^3\)$ in 2+1 D and than $O\(1/|\v q|^4\)$ in 3+1 D. For Coulomb-like potentials under consideration in 2+1 and 3+1 D the singularity of $A^{ext}_{0}(\v q)$ is $O\(1/| \vec q|\)$ and $O\(1/{\vec q}^2\)$, correspondingly.  In 1+1 D for the Coulomb potentials similar to (\ref{1.5a})   the singularity of $A^{ext}_{0}(\v q)$ is just a logarithmic one \cite{Davydov2017,Sveshnikov2017}, hence, $Q^{PT}_{VP}\equiv 0$ again.   Therefore for the  Coulomb potentials similar to (\ref{1.5a},\ref{1.6}) in all the three spatial dimensions there follows $Q^{PT}_{VP}\equiv 0$.

However, beyond the first-order PT and/or in the whole subcritical region $Z<Z_{cr,1}$, where in presence of negative discrete levels the dependence of $\vr_{VP}(\v r)$ on the external field cannot be described by the PT-series similar to (\ref{2.6}) any more, vanishing of $Q^{ren}_{VP}$ requires a sufficiently more detailed consideration.

\begin{figure}
\center
\includegraphics[scale=0.28]{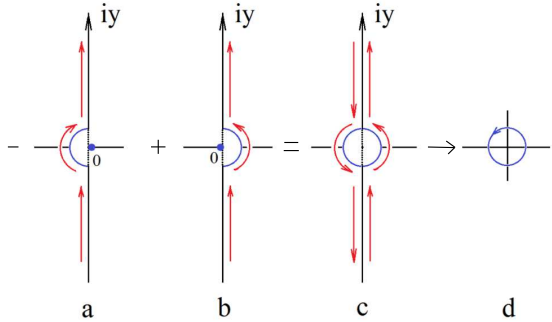} \\
\caption{\small (Color online) The picture of what happens with  the integral over the imaginary axis (\ref{2.11}), when the corresponding pole of the Green function moves along the real axis and  crosses zero point in the complex energy plane.}
\label{Residue}
\end{figure}

 In particular, the direct check confirms that the contribution of $\vr^{(3+)}_{VP}$ to  $Q^{ren}_{VP}$ for $Z<Z_{cr,1}$ vanishes too. In 1+1 D  this statement can be justified   purely analytically~\cite{Davydov2017}, while in 2+1 D due to complexity of  expressions, entering $\r^{(3+)}_{VP,|m_j|}(r)$, it requires a special combination of analytical and numerical methods (see Ref.~\cite{Davydov2018a}, App.B). For 3+1 D this combination is extended with minimal changes, since the structure of partial Green functions in two- and three-dimensional cases is actually the same up to additional factor $1/r$ and replacement $m_j \to k$. Moreover, it suffices to verify that  $Q_{VP}^{ren}=0$ not for the whole subcritical region, but only in absence of negative discrete levels. In presence of the latters,  vanishing of the  total VP-charge for $Z<Z_{cr,1}$ can be figured out by means of the  model-independent arguments, which are based on the initial  expression for the vacuum density (\ref{3.1}). It follows from (\ref{3.1}) that the change of $Q_{VP}^{ren}$ is possible only for  $Z>Z_{cr,1}$,  when the  discrete levels attain the lower continuum.   One of the possible correct ways to prove this statement is based on the rigorous analysis of the  behavior of the integral over the imaginary axis
 \beq \label{2.11}
 I_k(r)={1 \over 2\pi i}\,  \int\limits_{-i \inf}^{+i \inf} \! d\e\, \mathrm{Tr}G_k(r,r;\e) \ ,
 \eeq
 which enters the initial expression (\ref{3.14})  for $\vr_{VP,k}(r) $, under such infinitesimal variation of the external source, when the initially positive, infinitely close to zero level $\e_{n,k}$ becomes negative. Then the corresponding pole of the Green function undergoes an infinitesimal displacement along the real axis and also crosses the zero point, what yields the change in $I_k(r)$, equal to the residue at $\e=0$, namely
\beq \label{2.12}
 \D I_k(r)=\left. -|\p_{n,k}(r)|^2 \right|_{\e_{n,k}=0} \ .
 \eeq
For details see Fig.\ref{Residue} and take account of the definition of the Green function (\ref{3.3}).

So, by accounting for the degeneracy of levels in the spherically-symmetric case, the contribution from $I_k(r)$ to VP-charge loses  $2|e||k|$, whenever there appears the next negative discrete level  $\e_{n,k}$. Until this negative level exists, this loss of charge is compensated by the corresponding term, entering the sum  over negative discrete levels in (\ref{3.14}). Moreover, this term provides the continuity of the VP-density, when the level $\e_{n,k}$ crosses zero point. However, as soon as this level dives into the lower continuum,  the total VP-charge loses exactly $2|e||k|$.

 It should be specially mentioned that the latter effect is essentially non-perturbative and so completely included in $\vr_{VP}^{(3+)}$, while  $\vr_{VP}^{PT}$ is here out of work. So in the case under consideration one finds that upon renormalization the VP-charge turns out to be  non-vanishing only for $Z>Z_{cr,1}$ due to the non-perturbative effects, caused by diving of discrete levels into the lower continuum, in accordance with Refs.~\cite{Greiner1985a,Plunien1986,Mohr1998,Greiner2012,Rafelski2016}.

Thus, the behavior of the renormalized via eq.(\ref{3.27}) VP-density in the non-perturbative region turns out to be just the one, which should be expected from general assumptions about the structure of the electron-positron vacuum for $Z> Z_{cr}$. This result serves as a reasonable reference point for the whole renormalization procedure, and being combined with minimal subtraction prescription,  eliminates the inevitable arbitrariness in isolating the divergent part in the initial expression for $\vr_{VP}(\v r)$. This circumstance will be used further by renormalization of the VP-energy.

A more detailed picture of the changes in $\vr^{ren}_{VP}(r)$ for  $Z>Z_{cr,1}$ turns out to be quite similar to that considered in Refs.~\cite{Greiner1985a,Plunien1986,Greiner2012, Rafelski2016} by means of U.Fano approach to auto-ionization processes in atomic physics \cite{Greiner2012,Fano1961}. The main result is that, when the discrete level  $\p_{\n}(\v r)$ crosses the border of the lower continuum, the change in the   VP-charge density equals to
\beq\label{3.28}
\D\vr_{VP}^{ren}(\v r)=-|e|\times |\p_{\n} (\v r)|^2 \ .
\eeq
Here it  should be noted first that the original approach \cite{Fano1961} deals directly with the change of density of states $n(\v r)$ and so the change in the induced charge density (\ref{3.28}) is  just a consequence of the basic relation $\vr(\v r)= -|e| n(\v r)$.  Furthermore, such a jump in VP-charge density occurs for each diving level with its unique set of quantum numbers $\{\n\}$. So in the present case each discrete $2|k|$-degenerated level upon diving into the lower continuum yields the jump of total VP-charge  equal to  $(-2|e||k|)$. The other quantities, including the lepton number, should reveal a similar behavior. Furthermore, the Fano approach contains also a number of approximations. Actually, the expression (\ref{3.28}) is exact only in the  vicinity of the corresponding $Z_{cr}$, what  is shown by a number of concrete examples in Refs.~\cite{Davydov2017,*Sveshnikov2017,
Davydov2018a,Sveshnikov2019a,Voronina2019a}. Therefore, the most correct way to find $\vr_{VP}^{ren}(\v r)$ for the entire range of  $Z$ under study is provided by the expression (\ref{3.27}),  supplied with subsequent direct check of the expected integer value of  $Q_{VP}^{ren}$.

\section{VP-energy in the non-perturbative approach}

The most consistent way to explore the spontaneous positron emission is to deal  with the VP-energy $ \E_{VP} $, since there are indeed the changes in $ \E_{VP} $, caused by discrete levels diving, which  are responsible for creation of vacuum positrons. Actually, $ \E_{VP} $ is nothing else but the Casimir vacuum energy for the electron-positron system~\cite{Plunien1986}. The starting expression for $ \E_{VP} $ is
 \beq
\label{3.290}
\E_{VP}=\frac{1}{2}\(\sum\limits_{\e_{n}<\e_{F}} \e_n  -   \sum\limits_{\e_{n}\geqslant \e_{F}} \e_n \) \ .
\eeq
The expression  (\ref{3.290}) is obtained from the Dirac hamiltonian, written in the form, which is  consistent with the Schwinger prescription for the current  (for details see, e.g., Ref.~\cite{Plunien1986}), and is defined up to a constant, depending on the choice of the energy reference point. Following the general prescription for the Casimir energy calculations ~\cite{Itzykson1980,Plunien1986}, the natural choice of the reference point for $ \E_{VP} $ is the free case $A_{ext} = 0 $, which in the present case must be combined with the circumstance that, unlike the purely photonic Casimir effect,   there exists also an infinite set of discrete Coulomb levels. To pick out  exclusively the interaction effects, it is therefore necessary to subtract in addition from each discrete level the mass of the free electron at rest.
In fact, such definition of $ \E_{VP} $ follows from the basic properties of QFT (see, e.g., the basic monographs~\cite{Bogoli'ubov1959, Bjorken1965}), which imply that the  in- and out- fields and states in any QFT-problem under question  should be subject of a special limiting procedure. The essence of the latter is that in any interacting system the asymptotically free in- and out- fields and states cannot be treated independently of dynamics. Actually, the free case should be defined   via pertinent  turning-off operation,  applied only to the final steps of calculation. In the case under consideration  it is  the limit $Q \to 0$. The specifics of the present problem is that for any infinitesimally small $Q$ the discrete spectrum of Coulomb states survives. So apart of continua, the free field limit in this case includes an infinite set of units, corresponding to discrete levels, concentrated in the infinitesimal vicinity of the condensation point, which must be subtracted in such a way, that preserves the physical content of the problem.

Thus, in the physically motivated form and in agreement with $\vr_{VP}(\v r)$, which is defined so that it automatically vanishes in the free case, the initial expression for VP-energy should be written as
\begin{multline}
\label{3.29}
\E_{VP}=\1/2 \(\sum\limits_{\e_n<\e_F} \e_n-\sum\limits_{\e_n \geqslant \e_F} \e_n + \sum\limits_{-1\leqslant \e_n<1} \! 1
\)_A \ - \\ - \ \1/2 \(\sum\limits_{\e_n \leqslant -1} \e_n-\sum\limits_{\e_n
	\geqslant 1} \e_n \)_0 \ ,
\end{multline}
where the label A denotes the  non-vanishing external field $A_{ext}$, while the label 0 corresponds to   $A_{ext}=0$.  Defined in such a way,  VP-energy vanishes by turning off the external field, while by turning on it contains only the interaction effects. Therefore  the expansion of $ \E_{VP} $ in (even) powers of the external field starts from $ O\(Q^2\) $.

Now let us extract  from (\ref{3.29}) separately the contributions from the discrete and continuous spectra for each value of angular quantum number $k$, and afterwards use for the difference of integrals over the continuous spectrum $ (\int d {\v q}\, \sqrt{q^2+1})_A-(\int d {\v q}\,\sqrt{q^2+1})_0 $ the well-known technique, which represents this difference in the form of an integral of the elastic scattering phase $ \d_k(q)$~\cite{Sveshnikov2017,Rajaraman1982,Sveshnikov1991,Jaffe2004,Grashin2022a}.
 The final answer for $ \E_{VP}(Z,R) $, written as a partial series over angular number $k$, reads~\cite{Grashin2022a}
\beq
\label{3.30}
\E_{VP}(Z,R)=\sum\limits_{k=1}^{\inf} \E_{VP,k}(Z,R) \ ,
\eeq
where
\begin{multline}
\label{3.31}
\E_{VP,k}(Z,R)  = k\, \(\frac{1}{\pi} \int\limits_0^{\inf} \!   \  \frac{q \, dq }{\sqrt{q^2+1}} \ \d_{tot}(k,q) \ + \right. \\ \left. + \ \sum\limits_{\pm}\sum\limits_{-1 \leqslant \e_{n,\pm k}<1} \(1-\e_{n,\pm k}\)\) \ .
\end{multline}
In (\ref{3.31}) $ \d_{tot}(k,q) $ is the total phase shift for the given values of the wavenumber $q$ and angular number $\pm k$, including the  contributions from the scattering states from both  continua and  both parities for the radial DC problem with the  hamiltonian (\ref{3.13}). In the contribution of the discrete spectrum to $\E_{VP,k}(Z,R)$ the additional sum $\sum_{\pm}$ takes  also account of both parities.

Such approach to evaluation of $ \E_{VP}(Z,R)$ turns out to be quite effective, since for the external potentials of the type (\ref{1.5a},\ref{1.6})  each partial VP-energy turns out to be finite without any special regularization. First, $ \d_{tot}(k,q) $ behaves both in IR and UV-limits of the $q$-variable much better, than each of the scattering phase shifts separately.  Namely,  $ \d_{tot}(k,q) $ is finite for $ q \to 0 $ and behaves like  $ O(1/q^3) $ for $ q \to \inf $, hence, the phase integral in (\ref{3.31}) is always convergent. Moreover, $ \d_{tot}(k,q) $ is by construction an even function of the external field.  Second, in the contribution of bound states  to $ \E_{VP,k} $  the condensation point $ \e_{n,\pm k} \to 1 $ turns out to be regular, because  $ 1-\e_{n,\pm k} \sim O(1/n^2) $ for $ n \to \inf $. The latter circumstance permits to avoid intermediate cutoff of the Coulomb asymptotics of the external potential for $ r \to \inf $, what significantly simplifies all the subsequent calculations.

In terms of $ \E_{VP}(Z,R)$ the divergence of the theory shows up in divergence of the partial series (\ref{3.30}).  The latter problem can be solved along the lines of Refs.~\cite{ Davydov2018b,Sveshnikov2019b}, which deals with the similar  expansion for $ \E_{VP} $ in 2+1 D. In the present 3+1 D case due to a lot of technical details this topic is considered separately in Ref.~\cite{Grashin2022a}. The main result is that, as expected from general grounds discussed above in terms of $\vr_{VP}(\v r)$ and in accordance  with similar results in 1+1 and 2+1 D cases ~\cite{Davydov2017,*Sveshnikov2017,
Davydov2018a,Davydov2018b,Sveshnikov2019a,Sveshnikov2019b},  the partial series (\ref{3.30})  for $\E_{VP}(Z,R)$ diverges quadratically in the leading $O(Q^2)$-order. So it requires  regularization and subsequent renormalization, although each partial $\E_{VP,k}(Z,R)$  is finite without any additional manipulations. Moreover, the  divergence of the partial series (\ref{3.30}) is formally the same as in 3+1 QED  for the  fermionic loop with two external lines.

The need in the renormalization via fermionic loop follows also from the analysis of the properties of $\vr_{VP}$, which shows that without such UV-renormalization the integral VP-charge cannot acquire the expected integer value in units of $(-2|e|)$. In fact, the properties of $\vr_{VP}$ play here the role of a controller, which provides the implementation of the required physical conditions for a correct description of VP-effects beyond the scope of PT, since the latters cannot be tracked via direct evaluation of $\E_{VP}$ by means of the initial relations (\ref{3.290}),(\ref{3.29}).

Thus, in the complete analogy with the renormalization of VP-density  (\ref{3.27}), we should pass to the renormalized VP-energy by means of  relations
\beq
\label{3.58}
\E^{ren}_{VP}(Z,R) =\sum\limits_{k=1} \E^{ren}_{VP,k}(Z,R) \ ,
\eeq
where
\beq
\label{3.59}
\E^{ren}_{VP,k}(Z,R)=\E_{VP,k}(Z,R)+ \z_k(R) Z^2 \ ,
\eeq
while the renormalization coefficients $\z_k(R)$ are defined in the following way
\beq
\label{3.60}
\z_k(R) = \lim\limits_{Z_0 \to 0}  \[{\E_{VP}^{PT}(Z_0)\,\d_{k,1}-\E_{VP,k}(Z_0) \over Z_0^2}\]_{R} \ . \eeq
The  essence of relations  (\ref{3.58}-\ref{3.60}) is to remove (for fixed  $R$\,!) the divergent $O(Q^2)$-components from the non-renormalized partial terms  $ \E_{VP,k}(Z,R) $ in the series (\ref{3.30}) and  replace them further by renormalized via fermionic loop  perturbative contribution to VP-energy $\E^{PT}_{VP}\,\d_{k,1}$. Such procedure provides simultaneously the convergence of the regulated this way partial series (\ref{3.58}) and the correct limit of $\E^{ren}_{VP}(Z,R)$  for $Q \to 0$ with fixed $R$.

Another way to achieve the renormalization prescription (\ref{3.58}-\ref{3.60}) can be based on the  Schwinger relation~\cite{Schwinger1949,*Schwinger1951,Sveshnikov2017,Plunien1986, Greiner2012} for the renormalized VP-quantities
\beq\label{3.61}
\d \E^{ren}_{VP}=\int \! \mathrm{d} \v r \ \vr^{ren}_{VP}(\v r)\, \d A_0^{ext}(\v r) +\d \E_N \ ,
\eeq
where $\E_N$ is responsible only for jumps in the VP-energy, caused by discrete levels crossing through the border of the lower continuum, and so is an essentially non-perturbative quantity, which  doesn't need any  renormalization. From (\ref{3.61}) there follows  that VP-energy and VP-density must be renormalized in a similar way by means of the same subtraction procedure.

The relation (\ref{3.61}) can be represented also in the partial form
\beq\label{3.62}
\d \E^{ren}_{VP,k}=\int\limits_0^{\inf} \! r^2\, dr \ \vr^{ren}_{VP,k}(r)\, \d A_0^{ext}(r) +\d \E_{N,k} \ ,
\eeq
from which there follows that the  convergence of partial series for  VP-density implies the convergence of  partial series for  VP-energy and vice versa.  $\E_{N,k}$ is always finite and, moreover, for any finite $Z$ vanishes for $k \geq k_{max}(Z)$, therefore doesn't influence the convergence of the partial series.

The perturbative term $\E^{PT}_{VP}\,\d_{k,1}$ is obtained from the general first-order Schwinger relation~\cite{Schwinger1949,*Schwinger1951}
\beq
\label{2.1}
\E^{PT}_{VP}=\frac{1}{2} \int d \v r\, \vr^{PT}_{VP}(\vec{r})\,A_{0}^{ext}(\vec{r}) \ ,
\eeq
where $\vr^{PT}_{VP}(\vec{r})$ is defined in (\ref{2.2}-\ref{2.4a}).
From (\ref{2.1}) and (\ref{2.2},\ref{2.3}) one finds
\beq
\label{2.13}
\E^{PT}_{VP}=\frac{1}{64 \pi^4}\, \int \! d \v q \ {\v q}^2\, \Pi_R(-{\v q}^2)\,  \Big| \int \! d \v r\, \mathrm{e}^{i \vec{q} \vec{r}}\,A_{0}^{ext}(\vec{r}) \Big|^2  \ .
\eeq
It should be noted that, since the function  $S(x)$, defined in (\ref{2.4},\ref{2.4a}), is strictly positive, the perturbative VP-energy is positive too.

In the spherically-symmetric case  with $A_{0}^{ext}(\vec{r})=A_0(r)$  the perturbative VP-term belongs to the $s$-channel, which gives the factor $\d_{k,1}$, and
\begin{multline}
\label{2.14}
\E^{PT}_{VP}=\frac{1}{\pi}\, \int\limits_0^{\inf} \! d q \  q^4\, \Pi_R(-q^2) \ \times \\ \times \  \( \int\limits_0^{\inf} \! r^2\, d r\, j_0(q r)\, A_{0}(r) \)^2  \ ,
\end{multline}
 whence for the sphere there follows
\beq
\label{2.15}
\E^{PT}_{VP,\,sphere}={ Q^2  \over 2 \pi R}\, \int\limits_0^{\inf} \! {d q \over q} \ S(q/m) \ J_{1/2}^2(q R) \ ,
\eeq
while for the ball one obtains
\beq
\label{2.16}
\E^{PT}_{VP,\,ball}={ 9\,  Q^2  \over 2 \pi R^3}\, \int\limits_0^{\inf} \! {d q \over q^3} \ S(q/m) \ J_{3/2}^2(q R) \ .
\eeq
For $R$ close to $R_{min}(Z)$, the integrals (\ref{2.15},\ref{2.16}) can be  calculated analytically, since for such $R$  there holds
\beq
\label{2.18}
m R \ll 1 \ .
\eeq
The latter condition is satisfied by the Coulomb source with relation (\ref{1.8})  up to $Z \sim 1000$ . In particular,
\beq\label{2.20}
\E^{(1)}_{VP,\,sphere}={ Q^2  \over 3 \pi R}\,  \[ \ln\( {1 \over 2 m R}\) - \g_E + {1 \over 6}\]
\eeq
for the sphere and
\beq
\label{2.21}
\E^{PT}_{VP,\,ball}={2\, Q^2 \over 5 \pi R}\,  \[ \ln\( {1 \over 2 m R}\) - \g_E +  {1 \over 5} \]
\eeq
for the ball (see Ref.~\cite{Plunien1986} for details).
It should be noted, however,  that the expressions (\ref{2.20},\ref{2.21}) are valid only under condition (\ref{2.18}), when  $\ln\( 1 / 2 m R \) \gg 1 $. In the latter case the ratio
\beq
\label{2.22}
\E^{PT}_{VP,\,ball}/\E^{PT}_{VP,\,sphere} \simeq 6/5 \
 \eeq
is just the same as for their classical electrostatic self-energies $ 3 Z^2 \a/5 R$ and $ Z^2 \a/2 R$.

For $R \gg R_{min}$ the condition (\ref{2.18}) fails and the evaluation of $\E^{PT}_{VP}(R)$ is performed  numerically. By means of existing nowadays soft- and hardware this task does not pose any problems.

\section{The results of calculations}

As in Ref.~\cite{Grashin2022a}, for greater clarity  of results  we restrict this presentation to the case of charged sphere (\ref{1.5a}) on the interval $0 < Z < 500$ with the numerical coefficient in eq. (\ref{1.8}) chosen as\,\footnote{With such a choice for a charged ball with $Z=170$ the lowest $1s_{1/2}$-level lies precisely at $\e_{1s}=-0.99999$. Furthermore, it is quite close to $1.23$, which is the most commonly used coefficient in heavy nuclei physics.}
\beq
\label{6.1}
R(Z)=1.228935\, (2.5\, Z)^{1/3} \ \hbox{fm} \ .
\eeq
On this interval of $Z$  the main contribution to VP-energy comes from the partial channels with $k=1\,,2\,,3$, in which a non-zero number of discrete levels has already dived into the lower continuum.

\begin{figure*}[t!]
\subfigure[]{
		\includegraphics[width=1\columnwidth]{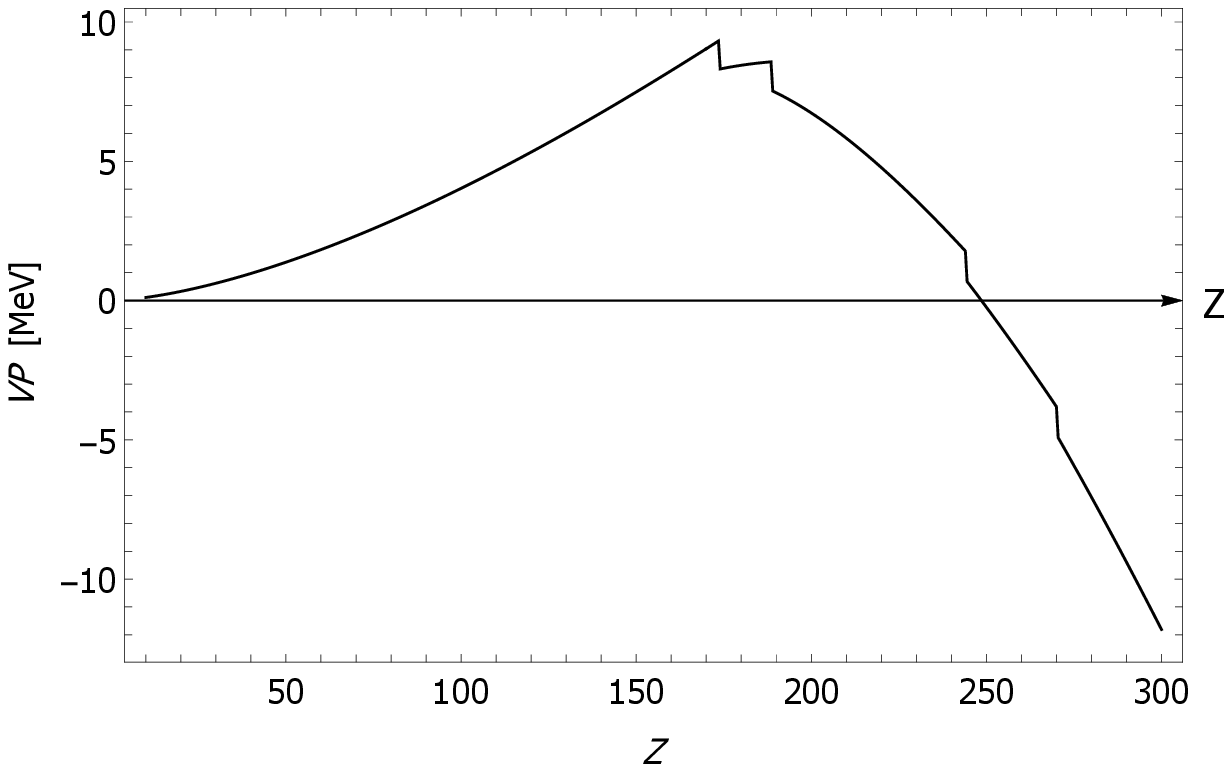}
}
\hfill
\subfigure[]{
		\includegraphics[width=1\columnwidth]{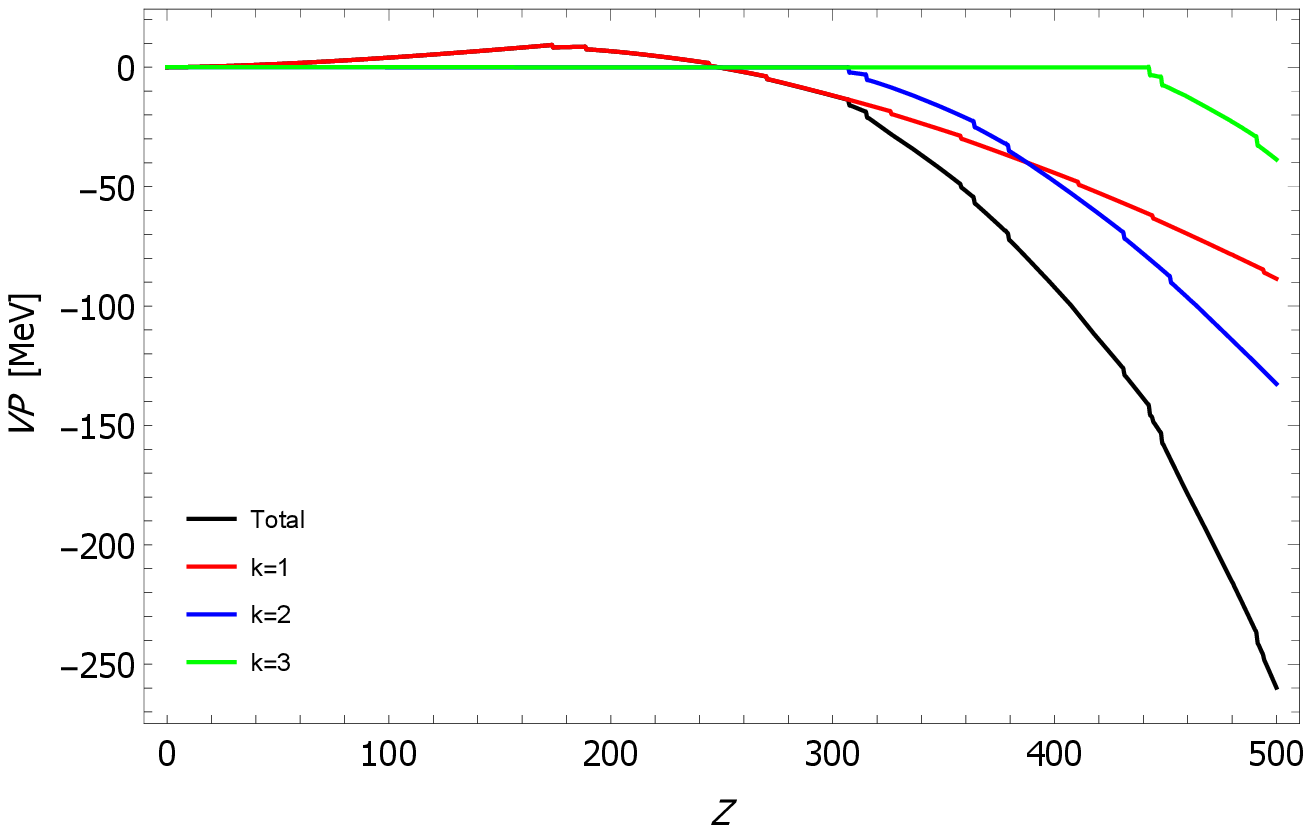}
}
\caption{(Color online) VP-energy for $R=R_{min}(Z)$: (a) $\E_{VP}^{ren}(Z)$ for $10 < Z < 300$; (b) partial $\E_{VP,k}^{ren}(Z)$ and $ \E_{VP,tot}^{ren}(Z)=\sum_{k=1}^{3}\E_{VP,k}^{ren}(Z) $ on the interval  $0 < Z < 500$ . }
	\label{VP(Z)}	
\end{figure*}

It would be instructive to start with  the  renormalized VP-energies  as functions of $Z$ with $R=R_{min}(Z)$, shown in Fig.\ref{VP(Z)}, and pertinent plots of $ \d_{tot}(k,q) $, shown in Figs.\ref{Phase(Z=300,R=Rmin)}, \ref{Phase(Z=500,R=Rmin)} (for details of calculations see Ref.~\cite{Grashin2022a}).
\begin{figure*}[bt!]
\subfigure[]{
		\includegraphics[width=1\columnwidth]{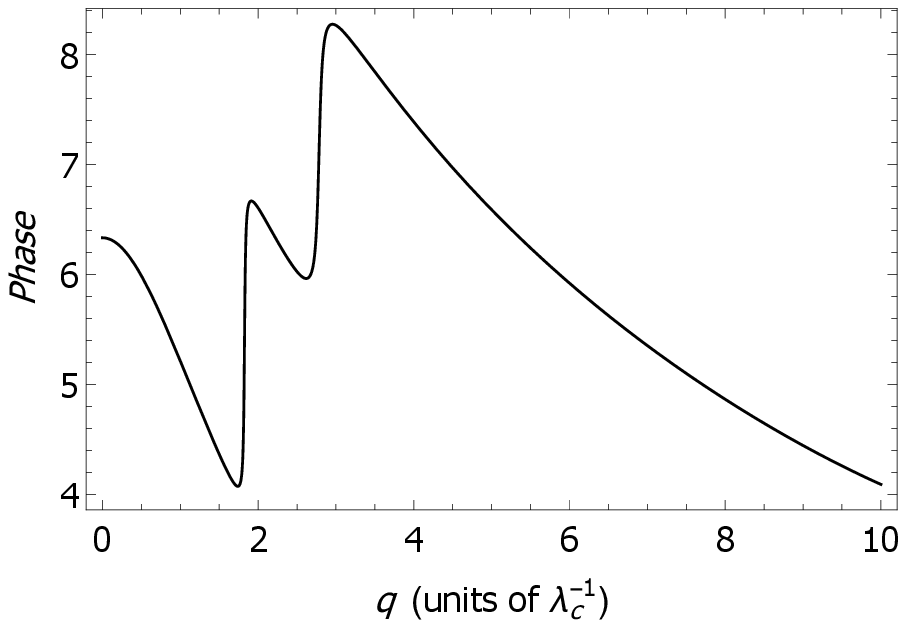}
}
\hfill
\subfigure[]{
		\includegraphics[width=1\columnwidth]{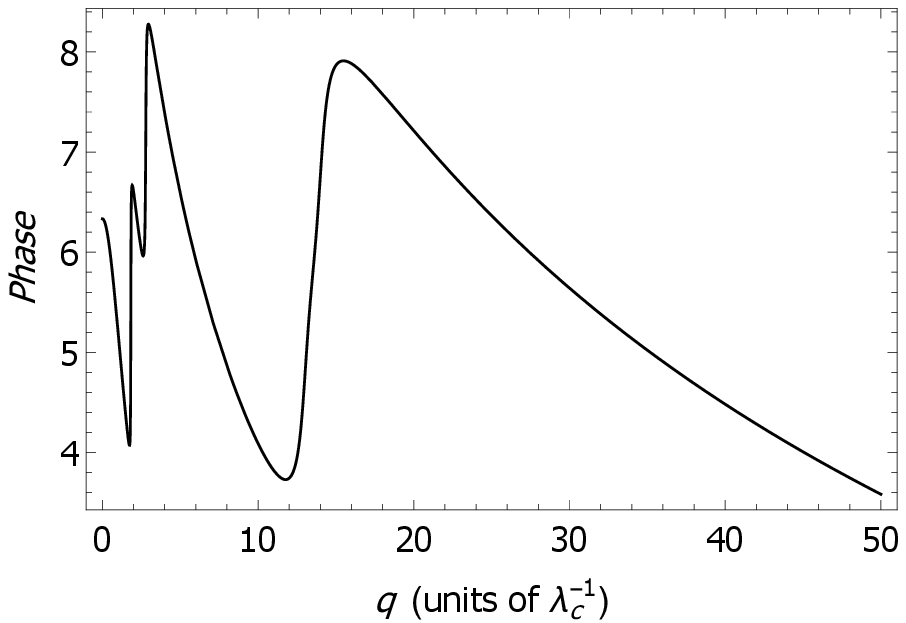}
}
\caption{$ \d_{tot}(1,q) $  for $Z=300$  and $R=R_{min}(Z)$ on different scales: (a)  for $0 < q < 10$; (b) for  $0 < q <50 $. }
	\label{Phase(Z=300,R=Rmin)}	
\end{figure*}
\begin{figure*}[t!]
\subfigure[]{
		\includegraphics[width=1\columnwidth]{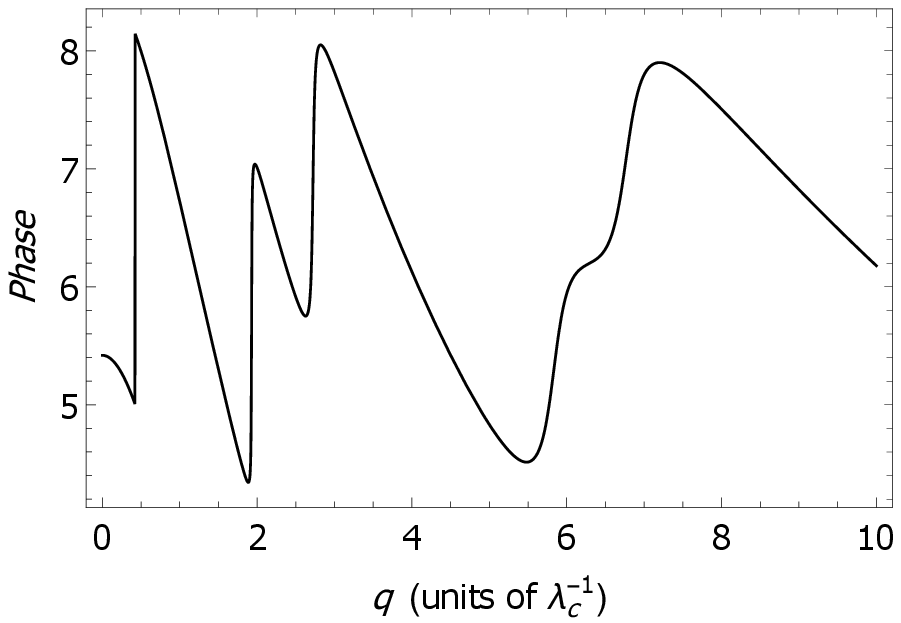}
}
\hfill
\subfigure[]{
		\includegraphics[width=1\columnwidth]{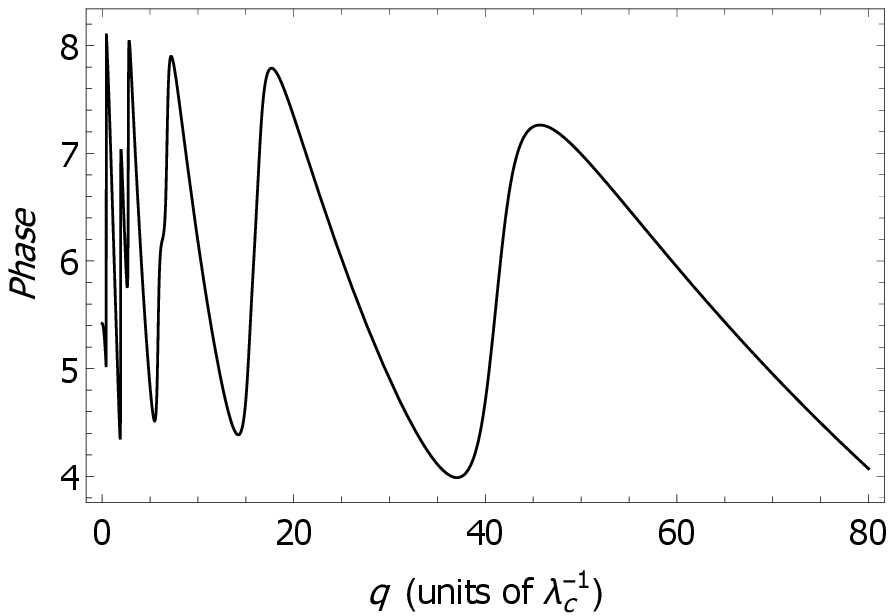}
}
\vfill
\subfigure[]{
		\includegraphics[width=1\columnwidth]{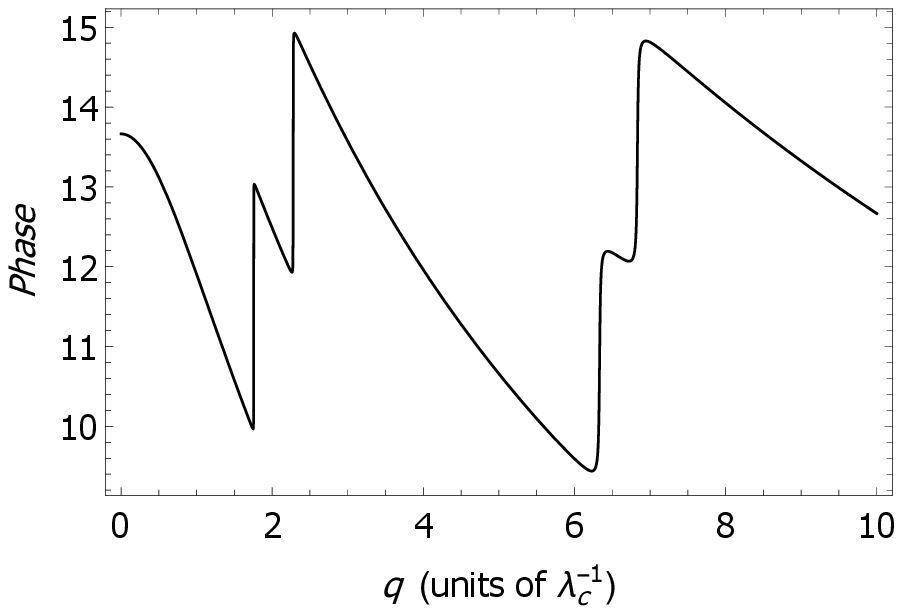}
}
\hfill
\subfigure[]{
		\includegraphics[width=1\columnwidth]{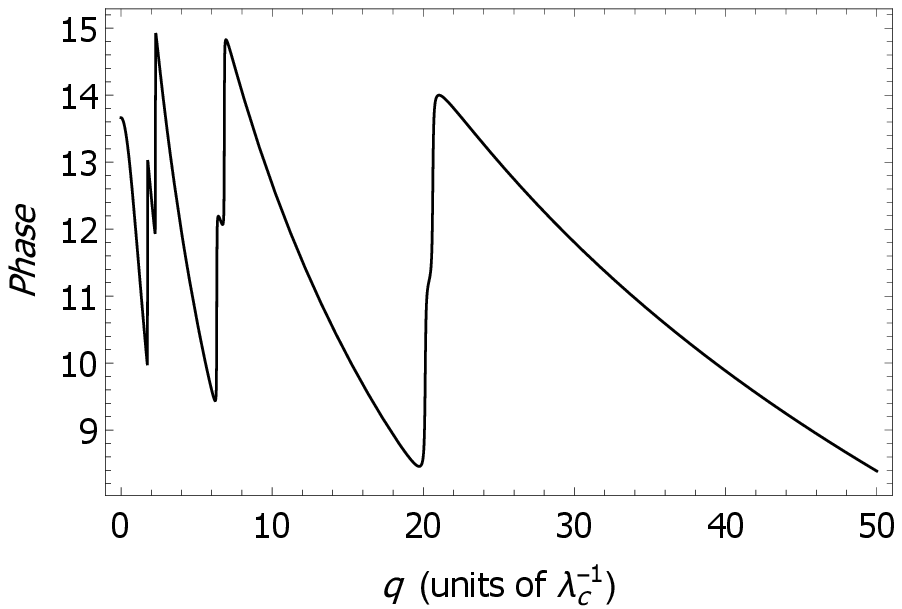}
}
\vfill
\subfigure[]{
		\includegraphics[width=1\columnwidth]{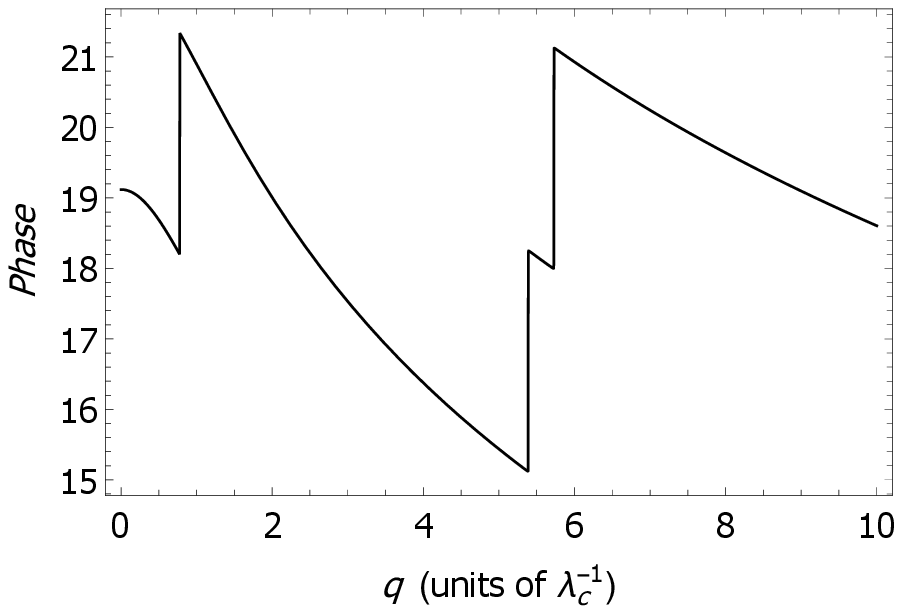}
}
\hfill
\subfigure[]{
		\includegraphics[width=1\columnwidth]{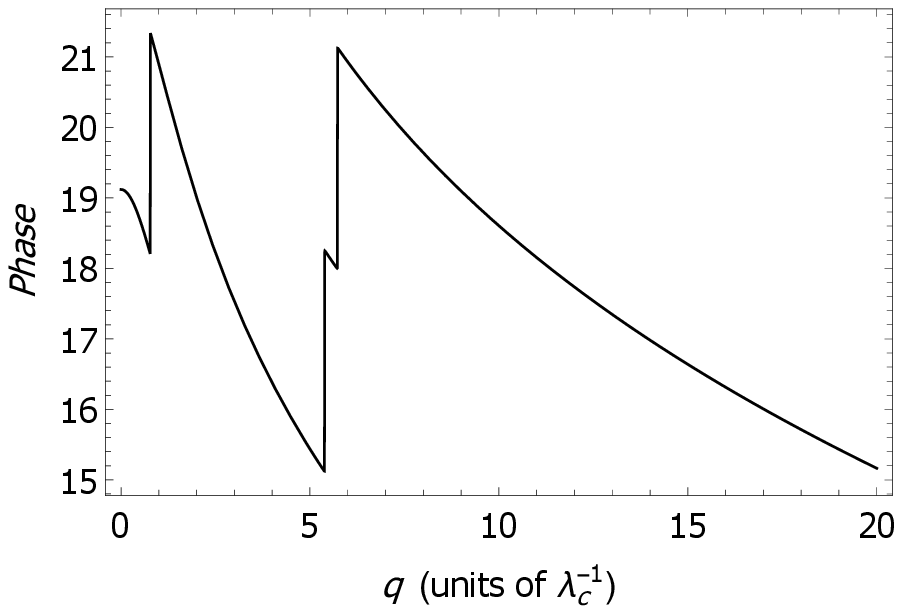}
}
\caption{$ \d_{tot}(k,q) $  for $Z=500$  and $R=R_{min}(Z)$ on different scales: (a,b) $ \d_{tot}(1,q) $  for $0 < q < 10$ and  $0 < q <80 $; (c,d) $ \d_{tot}(2,q) $  for $0 < q < 10$ and  $0 < q <50 $; (e,f) $ \d_{tot}(3,q) $  for $0 < q < 10$ and  $0 < q < 20 $. }
	\label{Phase(Z=500,R=Rmin)}	
\end{figure*}

The critical charges for both parities $(\pm k)$ are found in this case from the equations
\begin{multline}\label{5.9}
2\, z_1\,K_{2 i \eta_k}\(\sqrt{8\,Q R}\)\,J_{\pm} \ \mp  \\ \mp \ \[\sqrt{2\,Q R}\,\(K_{1+2\, i \eta_k}\(\sqrt{8\,Q R}\) + K_{1-2 i \eta_k}\(\sqrt{8\,Q R}\)\)  \pm \right. \\ \left. \pm \ 2\,k\,K_{2 i \eta_k}\(\sqrt{8\,Q R}\)\]\,J_{\mp}=0 \ ,
\end{multline}
where $K_{\n}(z)$ is the McDonald function,
 \beq\label{4.29}
\eta_k=\sqrt{Q^2-k^2} \ ,
\eeq
and
\beq\label{5.10}
 z_1=\sqrt{Q^2-2\, Q R} \ , \quad J_{\pm}=J_{k \pm 1/2}(z_1) \  .
\eeq
The meaning of eqs. (\ref{5.9}-\ref{5.10}) is twofold. First,   by solving these eqs. with respect to $Z$ and $R=R_{min}(Z)$ one finds the standard set of critical charges $Z_{cr,i}$ for the case of charged sphere. Second, for fixed $Z$ one obtains the set of critical radii $R_{cr,i}(Z)$, for which the levels attain the threshold of the lower continuum.

For $Z=300$ there are four lowest levels in the $s$-channel, which have been already dived into the lower continuum. The dived levels  always group quite naturally into pairs, containing states of different parity. In this case these are the pairs $\{1s_{1/2}\, , 2p_{1/2}\}$ and $\{2s_{1/2}\, ,  3p_{1/2}\}$ with $Z_{cr} \simeq 173.6\,, 188.5\, , 244.3\, , 270.5$, correspondingly. So for $Z=300$ it is only the $s$-channel, where $\d_{tot}(k,q)$ undergoes corresponding resonant jumps by $\pi$. The pertinent curves of $\d_{tot}(1,q)$ are shown in Fig.\ref{Phase(Z=300,R=Rmin)}. In particular, the two first low-energy narrow jumps   correspond to resonances, which  are caused by diving of $2s_{1/2}$ and  $3p_{1/2}$, what happens quite close to $Z=300$. At the same time, the jumps caused by diving of $1s_{1/2}$ and  $2p_{1/2}$ have been already gradually smoothed and almost merged together, therefore look like one big $2\,\pi$-jump, which is already significantly shifted to the region of larger $q$. In the other channels with $k \geq 2$ there are no dived levels for such $Z$. Therefore $\d_{tot}(k,q)$ in these channels looks like a decreasing function, whose behavior  roughly resembles the one of WKB-phase up to oscillating tail~\cite{Grashin2022a}. Furthermore, the contribution from these channels to VP-energy is negligibly small compared to those with diving.

For $Z=500$ there are already 3 separate $k$-channels with dived levels, namely  $k=1\, , 2\, , 3$.  For $k=1$ these are four pairs  $\{1s_{1/2}\,, 2p_{1/2}\}\,, \dots\,, \{4s_{1/2}\,, 5p_{1/2}\}$, with the last unpaired $5s_{1/2}$, corresponding to $Z_{cr} \simeq 494.3$. For $k=2$ it is the set $\{2p_{3/2}\,, 3d_{3/2}\}\,, \{3p_{3/2}\,, 4d_{3/2}\}\,, \{4p_{3/2}\,, 5d_{3/2}\}$, while for $k=3$ one has the pair $\{3d_{5/2}\,, 4f_{5/2}\}$ and the third unpaired $4d_{5/2}\, \(Z_{cr}\simeq 491.3\)$. The behavior  of total phases in these channels is shown in Fig.\ref{Phase(Z=500,R=Rmin)} and  differs only by the number of resonant jumps from the $s$-channel for $Z=300$, therefore  in the appearance of curves. For the $s$-channel the total number of dived levels equals to 9 and so  $\d_{tot}(1,q)$ undergoes the corresponding number of jumps by $\pi$,  which at small $k$ practically overlap each other. At the same time, for $k=2$ and $k=3$ all the jumps are quite clearly pronounced, transparent, and lie in one-to-one correspondence with the sequence of dived levels.

It should be remarked, however, that each $Z_{cr,i}$, given above, corresponds to the case of external Coulomb source with charge $Z=Z_{cr,i}$ and radius $R_i=R_{min}(Z_{cr,i})$. Hence, this is not exactly  the case under consideration with fixed $Z=300$ or $Z=500$ and $R=R_{min}(Z)$, rather it is  a qualitative picture of what happens with the dived levels in this case. They are absent in the discrete spectrum and show up as positronic resonances, the rough  disposition of which can be  understood as a result of diving of corresponding levels at $Z_{cr,i}$. But their exact positions on the $q$-axis can be found only via thorough restoration of the form of $ \d_{tot}(k,q)$.

\begin{figure*}[bt!]
\subfigure[]{
		\includegraphics[width=1.5\columnwidth]{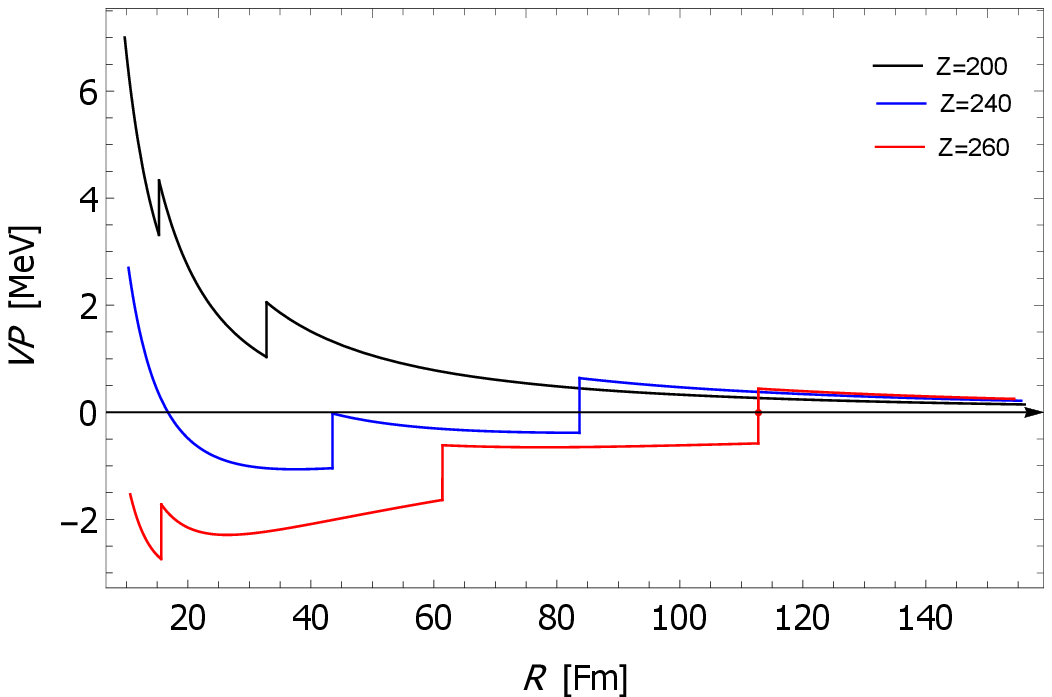}
}
\vfill
\subfigure[]{
		\includegraphics[width=1.5\columnwidth]{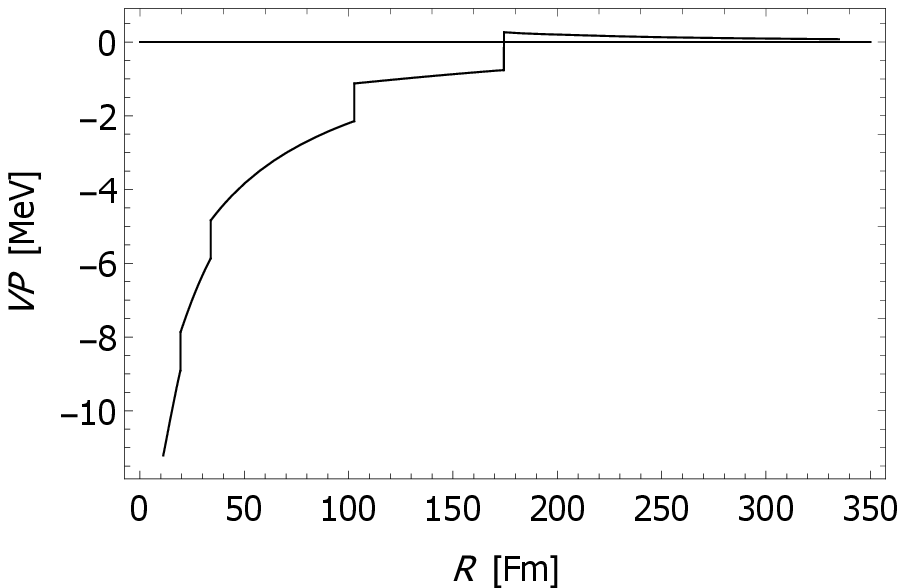}
}
\caption{(Color online) The behavior of  $ \E_{VP}^{ren}(Z,R)$ in the $s$-channel, which for such $Z$  is the dominating one and contains more than $99 \%$ of contributions from all the VP-effects under question  for: (a) $Z=200\, , 240\,,260$; (b) $Z=300$ .  }
	\label{VP200-300(R)}	
\end{figure*}

\begin{figure*}[t!]
\subfigure[]{
		\includegraphics[width=1\columnwidth]{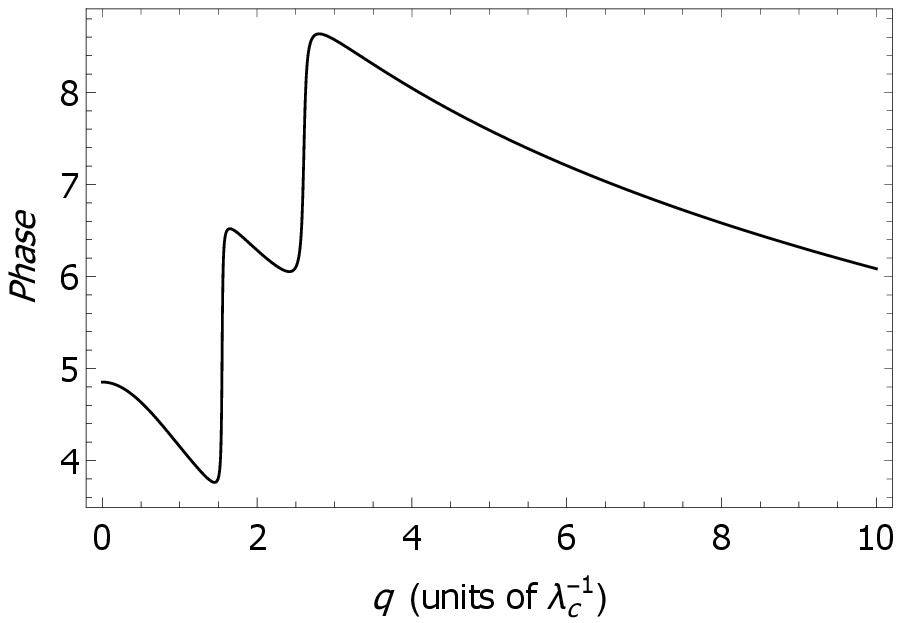}
}
\hfill
\subfigure[]{
		\includegraphics[width=1\columnwidth]{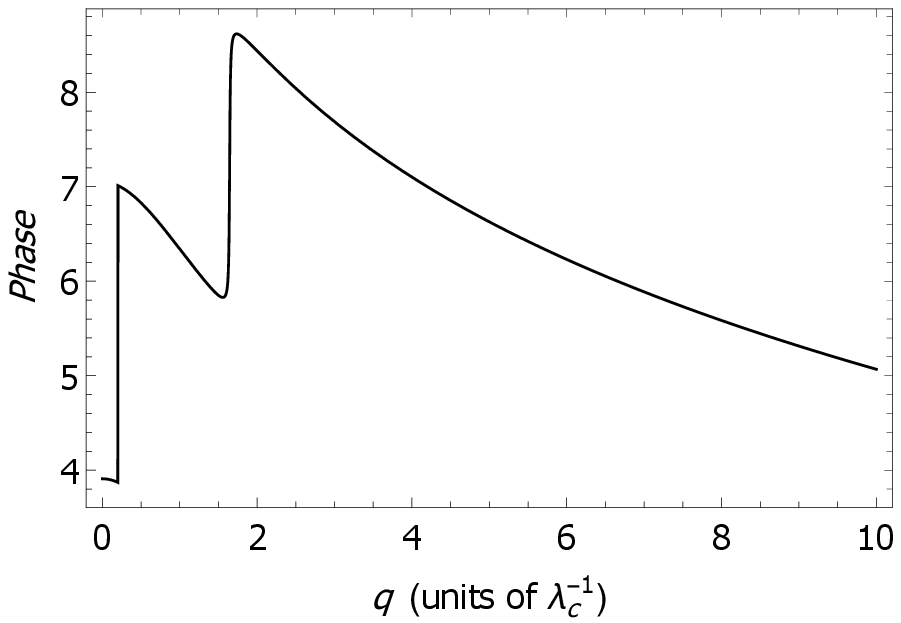}
}
\vfill
\subfigure[]{
		\includegraphics[width=1\columnwidth]{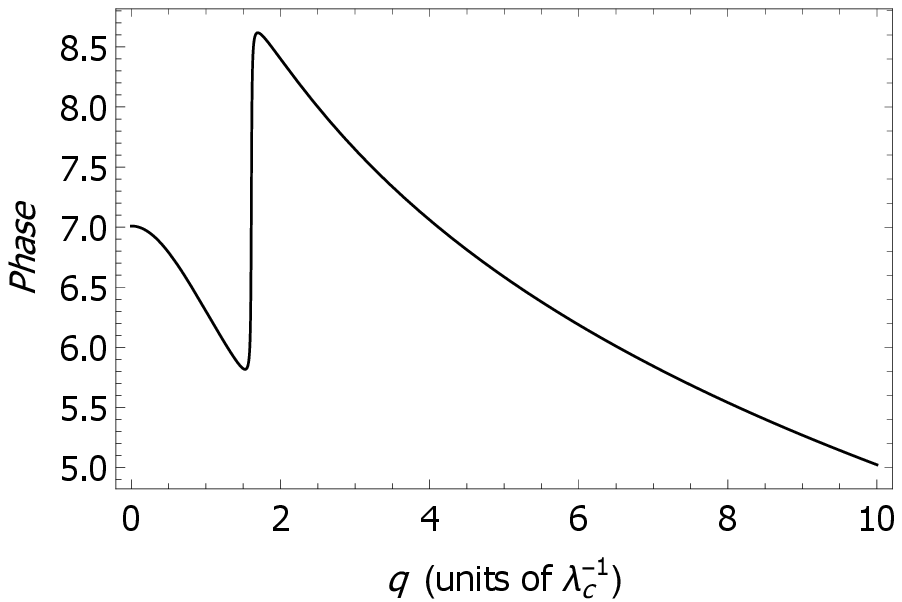}
}
\hfill
\subfigure[]{
		\includegraphics[width=1\columnwidth]{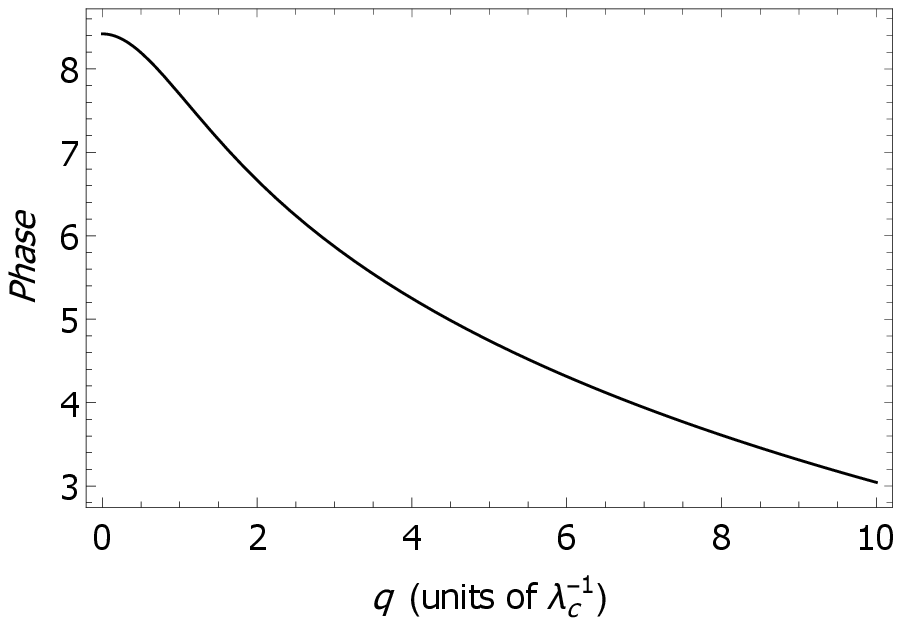}
}
\vfill
\subfigure[]{
		\includegraphics[width=1\columnwidth]{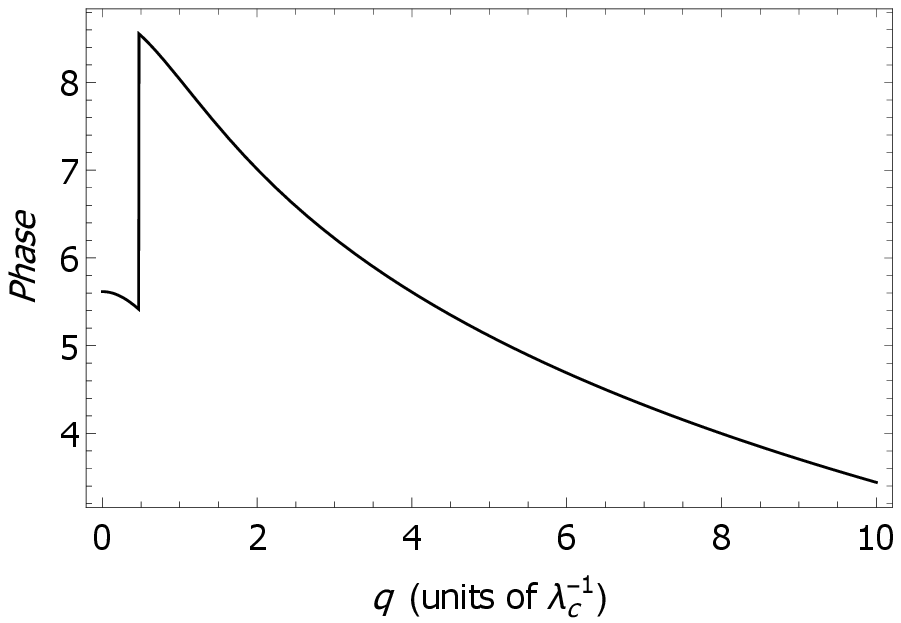}
}
\hfill
\subfigure[]{
		\includegraphics[width=1\columnwidth]{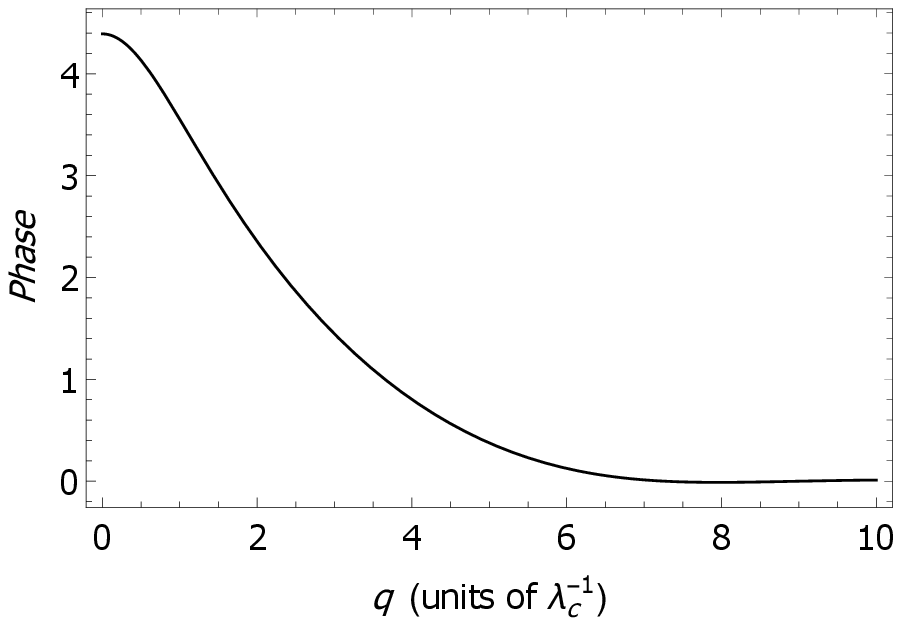}
}
\caption{$ \d_{tot}(1,q) $ for $Z=200$  and (a) $R=R_{min}(Z) \simeq 0.02525 $; (b) $R=1,55 R_{min}(Z) \simeq 0.03914$; (c)  $R=1,58 R_{min}(Z) \simeq 0.03989$; (d) $R=3 R_{min}(Z) \simeq 0.0757$;  (e) $R=3.5 R_{min}(Z) \simeq 0.08837$; (f) $R=20 R_{min}(Z) \simeq 0.505 $ .}
	\label{Phase(Z=200,R)}	
\end{figure*}

\begin{figure*}[t!]
\subfigure[]{
		\includegraphics[width=1.5\columnwidth]{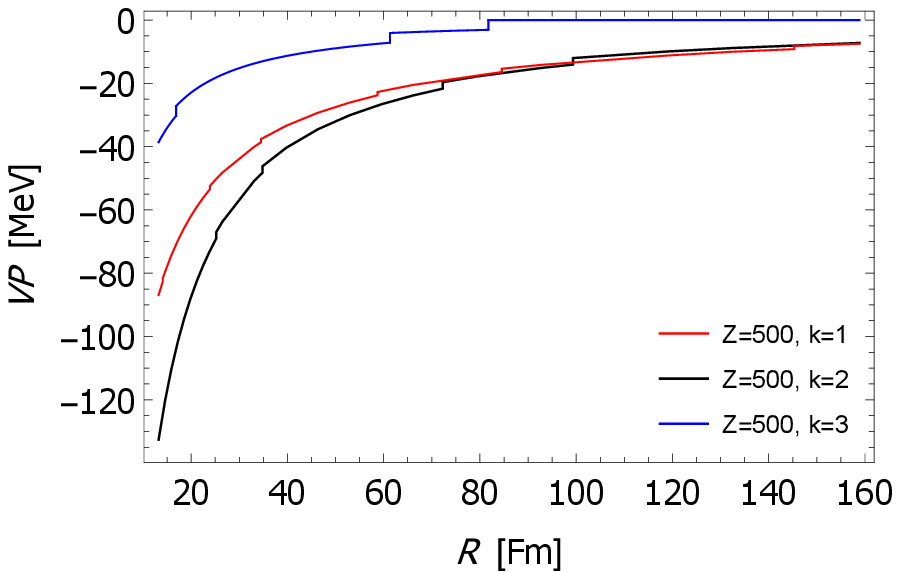}
}
\hfill
\subfigure[]{
		\includegraphics[width=1.5\columnwidth]{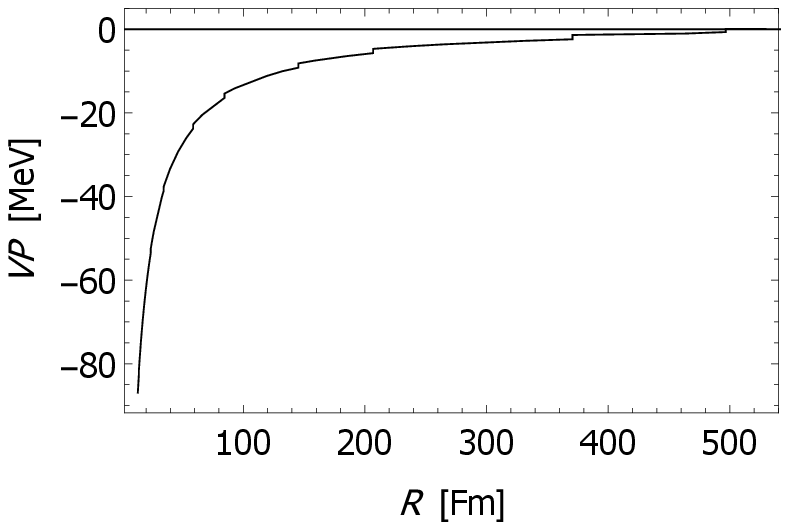}
}
\caption{(Color online) Partial  $ \E_{VP,k}^{ren}(R)$ for $Z=500$: (a) $k=1\,,2\,,3$ coupled together on the interval $0< R < 160$; (b) $k=1$ on a much larger interval $0< R < 550$ including all the jumps, caused by dived levels, and the asymptotical tail. }
	\label{VP500(R)}	
\end{figure*}

Now let us turn to the study of behavior of VP-energy as a function of $R$ for a representative set of supercritical $Z$. The pertinent plots are shown in Figs.~\ref{VP200-300(R)}-\ref{VP500(R)}. Each curve is given  on the interval $R_{min}(Z) < R < R_{max}$, where $R_{max} \sim 1$ is chosen in such a way that it corresponds to the $O(1/R)$-asymptotics of $ \E_{VP}^{ren}(Z,R)$. The latter for the $s$-channel with $k=1$ appears as a result of competition of contributions from $ \E_{VP}^{PT}(R)$, defined by the integral (\ref{2.15}), and from the other terms in $ \E_{VP,1}^{ren}(Z,R)$. For $Z=200-300$ this competition yields the  positive asymptotics, whereas for $Z=500$ it changes the sign due to increasing role of the non-perturbative component in $ \E_{VP,1}^{ren}(500,R)$.  However, in any case it is $O(1/R)$. For $Z=500$ in channels with $k=2\, , 3$ the perturbative contribution is absent and the asymptotics is strictly negative. It should be noted also that in the $s$-channel the study of large $R$-asymptotics requires indeed the use of the exact integral representation of  $ \E_{VP}^{PT}(R)$, given in  (\ref{2.15}), instead of the approximate analytic expression  (\ref{2.20}), since in this region the condition (\ref{2.18}) apparently fails. At the same time, at the left end for $R=R_{min}(Z)$ the values of $ \E_{VP}^{ren}(Z,R)$ coincide with those from Fig.\ref{VP(Z)}.

In Fig.\ref{VP200-300(R)}(a) there are shown the curves of  $ \E_{VP}^{ren}(Z,R)$ for $Z=200\, , 240\, , 260$, which reveal the same basic property. Namely, their curvatures are positive, and so they are bulge down. To the contrary, the curvature of $ \E_{VP}^{ren}(Z,R)$ for $Z=300$, which is intentionally shown in the separate Fig.\ref{VP200-300(R)}(b), is negative and the curve of $ \E_{VP}^{ren}(300,R)$ is bulging upwards.

For $Z=200$ with $R_{min}=0.02525$, that means $\simeq 9.7$ fm, there is only one pair of the lowest $s$-levels, namely $\{1s_{1/2}\, , 2p_{1/2}\}$, which dive into the lower continuum, now in reverse order compared to the dependence on $Z$. The most representative picture of diving in this case is given by the behavior of $\d_{tot}(1,q)$, which is shown in Fig.\ref{Phase(Z=200,R)}. First, with increasing $R$ from $R_{min}$ to $R_{max}$, it is the $2p_{1/2}$-level, which dives first at $R_{cr,1}=1.57062 R_{min}=0.04397$, whereas  $1s_{1/2}$ dives much further at $R_{cr,2}=3.36193 R_{min}=0.08489$. Such a difference in $R_{cr,i}$ between $1s_{1/2}$ and $2p_{1/2}$  appears, because $Z_{cr,2} \simeq 188.5$ for $2p_{1/2}$ lies much closer to $Z=200$ than $Z_{cr,1} \simeq 173.6$ for $1s_{1/2}$.

The behavior of $ \E_{VP}^{ren}(Z,R)$ for $Z=240\,,260$ differs from $Z=200$ only in the number of dived levels and in the arrangement of curves in Fig.\ref{VP200-300(R)}(a), in accordance with their starting values for $R=R_{min}(Z)$. The curve $ \E_{VP}^{ren}(200,R)$ is the highest one with $ \E_{VP}^{ren}(200,R_{min}(200))\simeq 13.7$, $ \E_{VP}^{ren}(240,R)$ with $ \E_{VP}^{ren}(240,R_{min}(240))\simeq 5.279$ takes  the middle position, whereas the curve $ \E_{VP}^{ren}(260,R)$ is the lowest one with $ \E_{VP}^{ren}(260,R_{min}(260))\simeq -3.0001$.

In contrast to $Z=200\, , 240\, , 260$, the behavior of $ \E_{VP}^{ren}(300,R)$ is principally different. First, it lies remarkably lower, since the starting value for $R=R_{min}(Z)$ is in this case sufficiently more negative $ \E_{VP}^{ren}(300,R_{min}(300))\simeq -21.94 $\,\footnote{This value is slightly different from the one, which is given in Ref.~\cite{Grashin2022a}, since in the present case we use throughout the interval $R_{min}(Z) < R < R_{max}$   the exact integral representation (\ref{2.15}) of  $ \E_{VP}^{PT}(R)$ instead of the approximate analytic answer  (\ref{2.20}). For $R$ close to $R_{min}(Z)$ the difference between them is just a correction, which does not play any significant role in the study of $Z$-dependence in VP-energy, considered in Ref.~\cite{Grashin2022a}. Moreover, for sufficiently large $Z >300$ the main contribution to $ \E_{VP}^{ren}(Z)$ comes from channels with $k >1$. For the present analysis, however, it is quite important, since it is indeed the exact representation (\ref{2.15}), which  provides the correct evaluation of $ \E_{VP}^{ren}(Z,R)$ for   $R \gg R_{min}(Z)$.} and reveals the curvature, which is opposite to the previous ones. Moreover, with decreasing $R$ the decline  of $ \E_{VP}^{ren}(300,R)$ into the negative range increases by each subsequent level diving. Indeed the latter property makes the spontaneous emission  a visible effect  for this $Z$.
 With further growth of $Z$ the decline  of $ \E_{VP}^{ren}(Z,R)$ into the negative range with decreasing $R$ becomes more and more pronounced, as it is clearly seen in Fig.\ref{VP500(R)}. Moreover, there follows from Fig.\ref{VP500(R)}(a), that for $Z=500$ it is the $k=2$-channel, which yields  the main contribution to the total VP-energy decline, but not to the rate of spontaneous emission (see below).

\section{Spontaneous positron emission: what is wrong and what is possible}

Now --- having dealt with the general features of levels diving and the behavior of $ \E_{VP}^{ren}(Z,R)$ in the $R$-variable this way --- let us turn to what happens with spontaneous emission of positrons in such a picture. This exercise turns out to be quite informative, since in this  picture the behavior of $\E_{VP}(Z,R)$ simulates in a quite reasonable way the non-perturbative VP-effects in slow heavy ion collisions, caused by the discrete levels diving into the lower continuum.

First, let us mention that the general theory~\cite{Greiner1985a,Plunien1986,Greiner2012, Rafelski2016}, based on the  framework~\cite{Fano1961}, predicts that  after level diving there appears  a metastable state with lifetime $\sim 10^{-19}$ sec, which decays into  the spontaneous positron emission  accompanied with vacuum shells formation  according to  the Fano rule (\ref{3.28}).  An important point here is  that due to spherical symmetry of the source, during this process all the angular quantum numbers and parity of the dived level are preserved by the metastable state and further by positrons created. Furthermore, the spontaneous emission of positrons should be caused solely by VP-effects  without any other channels of energy transfer. The corresponding positron spectra have been calculated first in Refs.~\cite{Reinhardt1981,*Mueller1988,Ackad2008}, and explored quite recently with more details in Refs.~\cite{Popov2018,*Novak2018,*Maltsev2018,
Maltsev2019,*Maltsev2020}. These spectra demonstrate, in particular,  that  the emission of low-energy positrons should be strongly suppressed by the repulsive interaction with the nuclei, while at high energy the spectra fall off exponentially. Actually,  strong suppression of spontaneous emission at low energies is  nothing else but the direct consequence of the  well-known answer in Coulomb scattering  for the probability to find the scattering particles at zero relative distances
\beq
\label{6.4a}
|\P(0)|^2= { 2\, \pi \vk  \over v\, |\mathrm{e}^{2 \pi \vk}-1| } \ ,
\eeq
where $\vk=e_1 e_2/ \hbar v$ with $v$ being the relative velocity. In the case of repulsion for small $v$ the relation (\ref{6.4a}) implies
\beq
\label{6.4b}
|\P(0)|^2 \to  { 2 \pi e_1 e_2  \over \hbar\, v^2 }\, \mathrm{e}^{- 2 \pi e_1 e_2 /\hbar v}  \ ,
\eeq
and since the spontaneous positrons should be created in the vicinity of the Coulomb center,  there follows from (\ref{6.4b}) that the suppression of emission at low energies must be  very strong.

Proceeding further, it would be instructive to start with the process of  contraction, when $R$ varies from $R_{max}$ to $R_{min}$. This process simulates the approach of heavy ions to each other starting from large distances. In this case, the curves of $\E_{VP}(Z,R)$ in Fig.\ref{VP200-300(R)}(a) immediately show that for $Z=200$ and $Z=240$ the spontaneous emission is strictly forbidden. The reason is that although the rest mass of positrons can be created via negative jumps in  VP-energy at corresponding $R_{cr,i}$, which are exactly equal to $2 \times mc^2$ in accordance with two possible spin projections,   to create a real positron scattering state  it is not enough due to repulsion  between positrons and the Coulomb source.  To supply the emerging vacuum positrons with corresponding potential energy, an additional decrease of  $\E_{VP}^{ren}(Z,R)$ just after levels diving point is required. However, the curves of $\E_{VP}(Z,R)$  for $Z=200$ and $Z=240$ do not share such option. On the whole interval  $R_{min} < R < R_{max}$,  with the only exception of negative jumps at critical points $R_{cr,i}$, with decreasing $R$ they reveal a  constant growth\,\footnote{The only, but to a high degree fantastic, possibility in this case is the appearance of vacuum positrons at the spatial infinity in states at rest. And although common wisdom says that they must be born in the vicinity of the Coulomb source, such possibility cannot be excluded at once. In any case, however, such positrons cannot be detected.}.

The approximate starting point   of spontaneous emission is presented by the curve $\E_{VP}(260,R)$. In this case there exists an interval between  diving points of $2s_{1/2}$  at $R_{cr,1}=0.040595$ and $2p_{1/2}$  at $R_{cr,2}=0.158939$, where by decreasing $R$ the curve $\E_{VP}(260,R)$  first goes down, providing sufficient decline $\simeq 0.65$ MeV for the positron emission with an appreciable probability. So for $Z=260$ the spontaneous emission after $2p_{1/2}$ level diving becomes already possible, but its intensity cannot be sufficiently high.

Actually,  the start-up for spontaneous emission should be estimated as $Z^{\ast} \simeq 250-260$, since it will not happen until $\E_{VP}(Z,R_{min}(Z))$ becomes negative. But there follows from Fig.\ref{VP(Z)} that the latter cannot occur before $Z$ exceeds $250$. Moreover, this estimate for $Z^{\ast} $ turns out to be quite insensitive to the concrete model of the Coulomb source. The main reason is that this estimate is closely related to the condition $\E_{VP}(Z,R_{min}(Z)) \simeq 0$. At this point, the difference between the sphere and ball charge configurations is small, since for $R=R_{min}(Z)$ it is reliably confirmed in Ref.~\cite{Grashin2022a} that their VP-energies are related via ratio (\ref{2.22}). So in the positive range the curve of the ball VP-energy lies higher than the corresponding curve for the sphere, in the negative range they interchange,  whereas their intersection takes place indeed at the point $\E_{VP}(Z,R_{min}(Z)) \simeq 0$, i.e. for the same $Z$ up to corrections in inverse powers of $\ln\( 1 / 2 m R_{min}(Z)\)$.  At the same time, in slow heavy ion collisions it is well-known that the VP-effects at short internuclear distances, achieved in the monopole approximation, are in rather good agreement with exact two-center ones~\cite{Reinhardt1981,*Mueller1988,Tupitsyn2010,Maltsev2019,*Maltsev2020},  and lie always in between those for sphere and ball upon adjusting properly the coefficient in the relation (\ref{1.8}).

However, for $Z \simeq Z^{\ast}$ the intensity of spontaneous emission cannot be high enough  for detection on the conversion pairs background. Actually, the positron emission becomes a visible effect starting from $Z \sim 300$ and with further growth of $Z$ its intensity increases very rapidly, which opens up the possibility for unambiguous detection for such $Z$.

It happens because  the behavior of $ \E_{VP}^{ren}(300,R)$   is opposite to the previous ones (compare the appearance of curves in Fig.\ref{VP200-300(R)}(a) and Fig.\ref{VP200-300(R)}(b)). As a result, with decreasing $R$ the decay rate  of $ \E_{VP}^{ren}(300,R)$ into the negative range increases by each subsequent level diving and so supplies the positrons just born with sufficient amount of energy for emission in the vicinity of the Coulomb source. And it is indeed the latter property, which makes the spontaneous emission  a quite visible effect  for this $Z$.

 With further growth of $Z$ the decline  of $ \E_{VP}^{ren}(Z,R)$ into the negative range with decreasing $R$ by each subsequent levels diving becomes more and more pronounced, as it is clearly seen in Fig.\ref{VP500(R)}. Here it should be specially noted that, although for $Z=500$ it is already the $k=2$-channel, which yields the main contribution to the total VP-energy,  the rate of spontaneous emission in this channel is nevertheless slightly less compared to $s$-channel. The reason is the degeneracy of levels. In the $s$-channel there are possible only two opposite spin projections, while for $k=2$ the degeneracy of levels is 4. At the same time, there follows from Fig.\ref{VP500(R)}(a) that the decline of partial $ \E_{VP,2}^{ren}(500,R)$ for decreasing $R$ is less than twice the decline of $ \E_{VP,1}^{ren}(500,R)$. Therefore in the process of spontaneous emission the $s$-channel remains the dominant one, but due to additional contributions from channels with $k=2\,,3$ the total rate of emission becomes significantly more high and so opens up the possibility for unambiguous detection for such $Z$.

   Now let us turn now  to the inverse process of extension, when $R$ varies from $R_{min}$ to $R_{max}$. This process is equally important as contraction, since it is an essential component of any  heavy ions collision. In this case the jumps  in $ \E_{VP}^{ren}(Z,R)$ at $R_{cr,i}$ are positive and should be treated in the following way. For $Z < Z^{\ast}$ the creation of vacuum positrons is strictly forbidden. Therefore the metastable hole in the lower continuum, created after discrete level diving at the corresponding $R_{cr,i}$, remains unoccupied by a sea electron. In this case during the  process of extension  the  discrete level simply jumps out of the lower continuum by passing through $R_{cr,i}$ and returns to the valence zone without any other consequences.

At the same time, in the case of level diving with spontaneous emission for $Z > Z^{\ast}$,  the situation is quite different. First, during the initial process of contraction the corresponding amount of energy exceeding $2 k \times mc^2$ is transferred from  $ \E_{VP,k}^{ren}(Z,R)$ to positrons, emitted in the $k$-channel, which carry this amount away together with the other quantities including momentum, spin, charge, lepton number, etc. The hole in the lower continuum, created after discrete level diving at $R_{cr,i}$, is now occupied and transmutes into the  corresponding vacuum shell in accordance with the Fano rule (\ref{3.28}). Hence, when during the reverse process of extension this level  jumps out of the lower continuum, the vacuum shell collapses and instead of it there appears a valence electron.

To provide the validity of such a picture it would be instructive to consider the  total energy balance in the system. The total energy itself is given by
\beq\label{6.5}
\E_{tot}=\E_{S}  + \E_{v} + \E_{p} + \E_{VP}^{ren} \ .
\eeq
 In  the  expression (\ref{6.5}) the first term $\E_{S}$ is the energy of the Coulomb source, which in the present case of the charged sphere is nothing else, but its  repulsive self-energy
 \beq\label{6.6}
\E_{S}= Z^2 \a/ 2 R \ ,
\eeq
while in the case of heavy ions  it includes two contributions, responding for their kinetic energy and Coulomb interactions
\beq\label{6.7}
\E_{S}=\E_{kin} + \E_{Coul} \ .
\eeq
In  turn,  $\E_{v}$ and $\E_{p}$ are responsible for the energy of valence electrons and positrons emitted.

Within the standard problem statement in the in-state there are no valence electrons and positrons, only the Coulomb source in the initial position with corresponding VP-energy. In terms of heavy ion collisions it implies a set of in-coming bare nuclei at large spatial separations, for which $\E_{tot}=\E_{kin}$. For $Z < Z^{\ast}$ the spontaneous emission is absent and so the out-state is either the same or reduces  to elastic scattering in the case of heavy ions. At the same time, for $Z > Z^{\ast}$ during the reverse process of extension at $R_{cr,i}$ for the energy balance in the system one obtains (for brevity we consider the $s$-channel)
 \beq\label{6.8}
\D \E_{VP}^{ren}=+ 2\, mc^2 \ ,
\eeq
while transmutation of the vacuum shell into the valence electron yields the following changes in $\E_{v}$
 \beq\label{6.9}
\D \E_{v}=+ 2\, mc^2 - 4\,mc^2= - 2\, mc^2 \ ,
\eeq
where $+ 2\, mc^2$ is the rest mass of electrons created, and it is taken into account that the valence electrons appear at the threshold of the lower continuum with the bound energy $- 2\, mc^2$. Thus, by passing through $R_{cr,i}$ during the extension process, for the total energy one finds
 \beq\label{6.10}
\D \E_{tot}=0 \ ,
\eeq
what confirms the correct status of such a picture.

As a result, in this case  the out-state includes the Coulomb source, but in a different position compared to the initial one, and a number of  valence electrons and spontaneous positrons. Moreover, the number of valence electrons created and positrons emitted must coincide in order to preserve the charge and lepton number.

\section{Concluding remarks}

First, it should be noted that the answer $Z^{\ast} \simeq 250-260$ for the start-up of spontaneous emission, given above, is just an estimate from below. Actually, $Z^{\ast}$ should be even larger, since this answer corresponds to a strongly underestimated  size of the Coulomb source via relation (\ref{6.1}) in comparison with the realistic two heavy ions supercritical configuration. Furthermore, there exist at least two more types of  corrections to $ \E_{VP}^{ren}(Z,R)$, which lead to a positive shift of $Z^{\ast}$. The first one is the screening of the Coulomb source by  vacuum shells, which is just a correction due to the  large spatial extension of the latters with asymptotics $\sim \exp\(-\sqrt{8\,Q\,r}\)$, but at the same time cannot be excluded at once. Another one is the self-energy (SE) contribution  due to virtual photon exchange, which in the context of essentially nonlinear VP-effects of super-criticality, caused by  fermionic loop,  is usually dropped. Near the lower continuum  SE shows up also as a perturbative correction   and so cannot seriously alter the results, presented above~\cite{Roenko2018}. Nevertheless,    SE is always positive and so leads to the same consequences for  $Z^{\ast}$. So the realistic estimate for $Z^{\ast}$ turns out to be even higher than  $ \simeq 250-260$. In any case, however, such $Z^{\ast}$ lies far beyond  the interval $170 \leqslant Z \leqslant 192 $, which is nowadays the main region of theoretical and experimental activity in heavy ion collisions aimed at the study of such VP-effects ~\cite{Popov2018,Maltsev2019,FAIR2009,Ter2015,MA2017169}.
The negative result of early investigations at GSI~\cite{Mueller1994} can be  at least partially explained by the last circumstance\,\footnote{We drop here all the questions concerning the time of level diving during heavy ion collisions. For more details on this topic see, e.g., Refs.~\cite{Rafelski2016, Greiner2012,Ruffini2010}.}.

       It should be also mentioned that  $Z^{\ast} \simeq 210 $  has already arisen by treating the spontaneous emission as a specific lepton pair creation~\cite{Grashin2022a,Krasnov2022,*Krasnov2022a}. The latter estimate  is significantly softer, but has been  obtained on the basis of more qualitative  arguments. The present one, however, is achieved in a quite rigorous way via direct evaluation of $ \E_{VP}^{ren}(Z,R)$ by means of well-founded and transparent approach.

       Second, it would be worth to figure out the  intimate, but crucial role of $\E_{VP}(Z,R)$ in the  energy balance in the system, especially for the case with spontaneous emission. For these purposes it would be instructive to consider directly a slow heavy ions collision with the in-state, containing  a set of in-coming bare nuclei, whereas in the final one there appears in addition  a number of emitted positrons and valence electrons. Actually, it is a typical inelastic process, which proceeds under condition of total energy conservation and in which the energy exchange between participants proceeds as follows. During the  contraction part of the  process $\E_{VP}(Z,R)$ decreases, and the corresponding part of the total decline of  $\E_{VP}(Z,R)$ is spent on the spontaneous positron emission, while  the rest of VP-decline leads to an increase in the kinetic energy of the nuclei due to $\D \E_{tot}=0$. This effect can be easily understood in terms of the   attractive VP-force, defined by the gradient of $ \E_{VP}^{ren}(Z,R)$ with respect to $R$
\beq
\label{7.1}
 F_{VP}=- \nabla_R\,\E_{VP}^{ren}(Z,R) \ ,
\eeq
which acts on the nuclei and leads to their additional acceleration. This force is by no means much smaller than  the Coulomb one, but the complete picture of spontaneous emission cannot be reproduced without taking account for this force.

   In particular, during the  extension part of the collision  $\E_{VP}(Z,R)$ permanently increases and $F_{VP}$ acts on the nuclei all the time as a braking force, excluding  the jumps at $R_{cr,i}$. So the kinetic energy of out-coming nuclei turns out to be less than that of in-coming. This loss of kinetic energy is  about few MeV and so is quite small compared to Coulomb repulsion between nuclei, but it is indeed this amount of energy, which provides the appearance of  the corresponding number of emitted positrons and valence electrons in the final state.

Third, in collisions with spontaneous emission the emitted positrons carry away the lepton number equal to $(-1)\times$their total number. Hence, the   corresponding amount of positive lepton number  should be left to the  VP-density, concentrated in vacuum shells. Otherwise, either the lepton number conservation  in such a processes must be broken, or the positron emission  prohibited. Therefore such collision should proceed via a specific intermediate state, which contains a spatially extended lepton number VP-density. Moreover,  it should be a real intermediate state, which appears with the first emitted positron and ceases  to exist with the last created valence electron. For slow heavy ion collisions its lifetime can be estimated as $\sim 10^{-19}-10^{-21}$ sec, and during this time the lepton number VP-density should exist as a real and, in principle,  measurable extended quantity.
So any reliable answer concerning the spontaneous positron emission --- either positive or negative --- is important for our understanding of the nature of the lepton number, since so far leptons show up as point-like particles with no indications on existence of any kind intrinsic structure (for a more detailed discussion of this topic see Ref.~\cite{Krasnov2022,*Krasnov2022a}).

Therefore, the reasonable conditions, under which the vacuum positron emission can be unambiguously detected on the nuclear conversion pairs background, should play an exceptional role  in slow  ions collisions, aimed at the search of such events. However, the estimates for $Z^{\ast}$, achieved above, show that for such conditions two heavy ion collisions are out of work. The gap between the highest achievable nowadays $Z=192$ for Cm+Cm colliding beams and the estimate from below $Z^{\ast} \simeq 250-260$ is to large for any kind special efforts in searching for the signal of spontaneous emission in such collisions.

  One possible way to access QED at a supercritical Coulomb field is to consider more complicated collisions, which are based on the regular polyhedron symmetry --- synchronized slow heavy ion beams move  from the vertices of the polyhedron towards its center. The simplest collision of such kind contains 4 beams, arranged as the mean lines of the tetrahedron with the intersection angle $\vt=2 \arcsin (\sqrt{2/3}) \simeq 109^{\small 0}\,28'$, akin to s-p hybridized tetrahedron carbon bonds. The next and probably the most preferable one is the 6-beam configuration, reproducing  6 semi-axes of the rectangular coordinate frame. The advantage of such many-beam collisions is evident. On the other hand, such collisions require serious additional efforts for their implementation. Therefore, an additional tasty candy is needed to arouse sufficient interest in such a project, and it is really possible, but this issue requires for a special discussion.

\section{Acknowledgments}

The authors are very indebted to Dr. Yu.S.Voronina, Dr. O.V.Pavlovsky   and P.A.Grashin from MSU Department of Physics and to Dr. A.S.Davydov from Kurchatov Center and Dr. A.A. Roenko from JINR (Dubna) for interest and helpful discussions.  This work has been supported in part by the RF Ministry of Sc. $\&$ Ed.  Scientific Research Program, projects No. 01-2014-63889, A16-116021760047-5, and by RFBR grant No. 14-02-01261. The research is carried out using the equipment of the shared research facilities of HPC computing resources at Moscow Lomonosov State University.

\bibliography{VP3DC}

\end{document}